\documentclass[useAMS,usenatbib,titlepage]{mn2e}
\usepackage{lscape,graphicx,amssymb,aas_macros,bm}
\usepackage{amsmath} 
\usepackage{indentfirst} 
\usepackage{pdflscape} 
\usepackage{placeins} 
\usepackage{nicefrac} 
\usepackage{afterpage} 
\usepackage{booktabs} 
\usepackage{color} 
\usepackage{float}

\newcommand{\lnhone}{\mbox{$\log\,N$(H\,{\scriptsize I})\ }}

\newcommand{\apg}{\:^{>}_{\sim}\:}
\newcommand{\apl}{\:^{<}_{\sim}\:}
\newcommand{\cmjj}{\mbox{${\rm cm^{-2}}$}}
\newcommand{\etal}{et al.}
\newcommand{\HI}{{\mbox{H\,{\scriptsize I}}}}

\newcommand{\kms}{\mbox{km\ s${^{-1}}$}}
\newcommand{\lya}{\mbox{${\rm Ly}\alpha$}}
\newcommand{\lyb}{\mbox{${\rm Ly}\beta$}}

\newcommand{\FeII}{{\mbox{Fe\,{\scriptsize II}}}}
\newcommand{\MgII}{{\mbox{Mg\,{\scriptsize II}}}}
\newcommand{\CaII}{{\mbox{Ca\,{\scriptsize II}}}}
\newcommand{\MgI}{{\mbox{Mg\,{\scriptsize I}}}}

\newcommand{\NII}{{\mbox{N\,{\scriptsize II}}}}
\newcommand{\NIII}{{\mbox{N\,{\scriptsize III}}}}
\newcommand{\NeIII}{{\mbox{Ne\,{\scriptsize III}}}}
\newcommand{\SiII}{{\mbox{Si\,{\scriptsize II}}}}
\newcommand{\CIII}{{\mbox{C\,{\scriptsize III}}}}
\newcommand{\CII}{{\mbox{C\,{\scriptsize II}}}}
\newcommand{\SiIII}{{\mbox{Si\,{\scriptsize III}}}}
\newcommand{\OVI}{{\mbox{O\,{\scriptsize VI}}}}
\newcommand{\NeVIII}{{\mbox{Ne\,{\scriptsize VIII}}}}
\newcommand{\OI}{{\mbox{O\,{\scriptsize I}}}}
\newcommand{\OII}{{\mbox{O\,{\scriptsize II}}}}
\newcommand{\OIII}{{\mbox{O\,{\scriptsize III}}}}
\newcommand{\OIV}{{\mbox{O\,{\scriptsize IV}}}}

\newcommand{\mstar}{{\mbox{$M_{\rm star}$}}}
\newcommand{\msun}{{\mbox{M$_{\odot}$}}}

\title[The Cosmic Ultraviolet Baryon Survey]{The Cosmic Ultraviolet Baryon Survey (CUBS) I.\ Overview and the diverse environments of Lyman limit systems at $\bm{z<1}$\thanks{Based on data gathered with the 6.5m Magellan Telescopes located at Las Campanas Observatory, ESO Telescopes at the Paranal Observatory, and the NASA/ESA {\it Hubble Space Telescope} operated by the Space Telescope Science Institute and the Association of Universities for Research in Astronomy, Inc., under NASA contract NAS 5-26555.}}

\author[Chen and the CUBS team]{Hsiao-Wen Chen$^{1}$\thanks{E-mail: hchen@oddjob.uchicago.edu}, 
Fakhri S.\ Zahedy$^{2}$, Erin Boettcher$^{1}$, Thomas M.\ Cooper$^{2}$, \newauthor Sean D.\ Johnson$^{2,3}$\thanks{Carnegie-Princeton Fellow}, Gwen C.\ Rudie$^{2}$, Mandy C.\ Chen$^{1}$, Gregory L.\ Walth$^{2}$, \newauthor  Sebastiano Cantalupo$^{4}$, Kathy L.\ Cooksey$^{5}$, Claude-Andr\'e Faucher-Gigu\`ere$^{6}$, \newauthor Jenny E.\ Greene$^{3}$, Sebastian Lopez$^{7}$, John S.\ Mulchaey$^{2}$, Steven V.\ Penton$^{8}$, \newauthor Patrick Petitjean$^{9}$, Mary E.\ Putman$^{10}$, Marc Rafelski$^{11,12}$, Michael Rauch$^{2}$,  \newauthor Joop Schaye$^{13}$, Robert A.\ Simcoe$^{14}$, and Benjamin J.\ Weiner$^{15}$ \\
$^{1}$Department of Astronomy \& Astrophysics, The University of Chicago, Chicago, IL 60637, USA \\
$^{2}$The Observatories of the Carnegie Institution for Science, 813 Santa Barbara Street, Pasadena, CA 91101, USA \\
$^{3}$Department of Astrophysics, Princeton University, Princeton, NJ 08544, USA \\
$^{4}$Department of Physics, ETH Wolfgang$-$Pauli$-$Strasse 27, 8093, CH-8093 Z\"urich, Switzerland \\
$^{5}$Department of Physics and Astronomy, University of Hawai'i at Hilo, Hilo, HI 96720, USA\\
$^{6}$Department of Physics \& Astronomy and Center for Interdisciplinary Exploration and Research in Astrophysics (CIERA),\\ Northwestern University, 1800 Sherman Ave, Evanston, IL 60201, USA \\ 
$^{7}$Departamento de Astronom\'ia, Universidad de Chile, Casilla 36-D, Santiago, Chile\\
$^{8}$Laboratory For Atmospheric and Space Physics, University of Colorado, Boulder, CO 80303, USA\\
$^{9}$Institut d’Astrophysique de Paris, CNRS-SU, UMR 7095, 98bis bd Arago, Paris F-75014, France \\
$^{10}$Department of Astronomy, Columbia University, New York, NY 10027, USA\\
$^{11}$Space Telescope Science Institute, Baltimore, MD 21218, USA \\
$^{12}$Department of Physics \& Astronomy, Johns Hopkins University, Baltimore, MD 21218, USA\\
$^{13}$Leiden Observatory, Leiden University, PO Box 9513, NL-2300 RA Leiden, the Netherlands\\
$^{14}$MIT-Kavli Institute for Astrophysics and Space Research; 77 Massachusetts Ave., Cambridge, MA 02139, USA \\
$^{15}$Steward Observatory, University of Arizona, Tucson, AZ 85721, USA \\
}

\begin{document}

\pagerange{\pageref{firstpage}--\pageref{lastpage}} \pubyear{2017}
\maketitle
\label{firstpage}
\begin{abstract}

We present initial results from the Cosmic Ultraviolet Baryon Survey
(CUBS).  CUBS is designed to map diffuse baryonic structures at
redshift $z\apl 1$ using absorption-line spectroscopy of 15 UV-bright
QSOs with matching deep galaxy survey data.  CUBS QSOs are selected
based on their NUV brightness to avoid biases against the presence of
intervening Lyman Limit Systems (LLSs) at $z_{\rm abs}<1$.  We report
five new LLSs of $\log\,N(\HI)/\cmjj\apg 17.2$ over a total redshift
survey pathlength of $\Delta\,z_{LL}=9.3$, and a number density
of $n(z)=0.43_{-0.18}^{+0.26}$.  Considering all absorbers with
$\log\,N(\HI)/\cmjj>16.5$ leads to $n(z)=1.08_{-0.25}^{+0.31}$ at
$z_{\rm abs} < 1$.  All LLSs exhibit a multi-component structure and
associated metal transitions from multiple ionization states such as
\CII, \CIII, \MgII, \SiII, \SiIII, and
\OVI\ absorption.
Differential chemical enrichment levels as well as ionization states
are directly observed across individual components in three LLSs.
We present deep galaxy survey data obtained using the VLT-MUSE
integral field spectrograph and the Magellan Telescopes, reaching
sensitivities necessary for detecting galaxies fainter than $0.1\,L_*$
at $d\apl 300$ physical kpc (pkpc) in all five fields.  A diverse
range of galaxy properties is seen around these LLSs, from a low-mass
dwarf galaxy pair, a co-rotating gaseous halo/disk, a star-forming
galaxy, a massive quiescent galaxy, to a galaxy group.  The closest
galaxies have projected distances ranging from $d=15$ to $72$ pkpc and
intrinsic luminosities from $\approx 0.01\,L_*$ to $\approx 3\,L_*$.
Our study shows that LLSs originate in a variety of
galaxy environments and trace gaseous structures with a broad range of metallicities.



\end{abstract}


\section{Introduction}
\label{section:introduction}

The circumgalactic medium (CGM) and intergalactic medium (IGM) contain
fuel for future star formation and a record of past feedback.  They
are uniquely sensitive to the physics of baryonic flows---one of the
principal missing ingredients in our understanding of galaxy evolution
(for reviews see e.g., Somerville \& Dav\'e 2015; Naab \& Ostriker
2017; Tumlinson \etal\ 2017).  While QSO absorption spectroscopy
provides a powerful tool for probing the diffuse gas phase in
intergalactic and circumgalactic space, a comprehensive study of the
CGM requires matching galaxy survey data.  Previous joint galaxy and
absorber studies have focused primarily on two disjoint epochs,
$z<0.4$ (e.g., Chen 2017; Kacprzak 2017; Tumlinson \etal\ 2017 for
recent reviews) and $z\approx 2$ (e.g., Steidel \etal\ 2010; Rudie
\etal\ 2012; Turner \etal\ 2014; Rudie \etal\ 2019).  However, as the
cosmic star formation rate density (SFRD) declines rapidly from
$z\approx 1.5$ to the present day, the CGM remains poorly constrained
over a significant fraction of cosmic history (see Burchett
\etal\ 2019 for recent effort in probing the warm-hot CGM through
observations of \NeVIII\ absorption).

Observations of the co-evolution of galaxies with their surrounding
gas complements the progress both in wide-field galaxy surveys and in
theoretical models of how galaxies form and evolve.  In particular,
state-of-the-art cosmological simulations, incorporating realistic
star-formation and feedback recipes, can both match the large-scale
statistical properties of galaxies and reproduce the observed
small-scale features (e.g., Vogelsberger \etal\ 2014; Hopkins
\etal\ 2014; Schaye \etal\ 2015; Wang \etal\ 2015; Dubois
\etal\ 2016).  But these models have fallen short in simultaneously
matching the spatial profiles of a wide range of heavy ions (such as
Mg$^+$, C$^{3+}$, O$^{5+}$) observed in the CGM where the majority of
the baryons reside (e.g., Hummels \etal\ 2013; Liang \etal\ 2016;
Oppenheimer \etal\ 2016; Nelson \etal\ 2018; Ji \etal\ 2019).  This
mismatch suggests that our understanding of the nature and effects of
gas inflows and outflows is still incomplete.  Identifying the missing
physics that governs the dynamical state of the CGM provides an
unparalleled constraint on the manner and mode of feedback in
galaxies.

To enable systematic studies of the diffuse CGM and IGM, we are
conducting the Cosmic Ultraviolet Baryon Survey (CUBS), which is a
large {\it Hubble Space Telescope} ({\it HST}) Cycle 25 General
Observer Program (GO-CUBS; ${\rm PID}=15163$; PI: Chen).  It is
designed to map the diffuse baryonic structures at $z\apl 1$, using
absorption-line spectroscopy of 15 UV bright QSOs with matching deep
galaxy survey data.  The primary goal of CUBS is to establish a legacy
galaxy and absorber sample to enable systematic studies of the
co-evolution of galaxies and their surrounding diffuse gas at a time
when the SFRD undergoes its most dramatic changes, thereby gaining key
insights into how galaxy growth is regulated by accretion and
outflows.  The CUBS program exploits the synergy between space-based
UV spectroscopy and ground-based wide-field surveys, as well as
optical echelle spectroscopy, for advancing a comprehensive
understanding of the cosmic evolution of baryonic structures.

Here we present initial results from the CUBS program, reporting five
new Lyman limit systems (LLSs) 
discovered at $z<1$ along the CUBS QSO sightlines.  In addition, we
present the galactic environment of these LLSs uncovered from an
ongoing galaxy survey in the CUBS fields, using the Magellan
Telescopes and the VLT Multi-Unit Spectroscopic Explorer (MUSE; Bacon
et al.\ 2010).  LLSs arise in optically thick gas with opacity
$\tau_{912}\apg 1$ to ionizing photons at rest-frame wavelength
$\approx 912$ \AA\ (or equivalently neutral hydrogen column density
$\log\,N(\HI)/\cmjj \apg 17.2$).
On cosmological
scales, the incidence of these optically-thick absorbers determines
the mean free path of ionizing photons and serves as a key ingredient
for computing the photoionization rate in the IGM (e.g., Rudie
\etal\ 2013; Faucher-Gigu\`ere 2020).  In individual galactic halos,
these absorbers are commonly seen at projected distances $d\apl 100$
kpc from known galaxies (e.g., Chen \etal\ 1998, 2001; Rudie
\etal\ 2012; Thom \etal\ 2012; Werk \etal\ 2014; Johnson \etal\ 2015;
Prochaska \etal\ 2017) with a mean covering fraction of
$\kappa_{\tau_{912}\ge 1}(d<100\,{\rm kpc})\apg 70$\% (e.g., Chen
\etal\ 2018).  The large scatter observed in both gas density and
metallicity of the absorbing gas (e.g., Zahedy \etal\ 2019; Lehner
\etal\ 2019) makes these absorbers a promising signpost of either
infalling clouds (e.g., Maller \& Bullock 2004; Faucher-Gigu\`ere \&
Kere\v{s} 2011; Fumagalli \etal\ 2011; van de Voort \etal\ 2012;
Afruni \etal\ 2019) or outflows (e.g., Faucher-Gigu\`ere \etal\ 2015,
2016) in galactic halos, or a combination thereof (e.g., Hafen
\etal\ 2017).  We examine these different scenarios based on the
galaxy environment revealed in the accompanying galaxy survey data.
%

The paper is organized as follows.  In Section 2, we describe the
design of the CUBS program and related spectroscopic observations.  We
describe the search and identification of LLSs along the CUBS QSO
sightlines in Section 3, and their galactic environments in Section 4.
In Section 5, we discuss the implications of our findings.  Throughout
the paper, we adopt a standard $\Lambda$ cosmology, $\Omega_M$ = 0.3
and $\Omega_\Lambda$=0.7 with a Hubble constant $H_{\rm 0} = 70\rm
\,km\,s^{-1}\,Mpc^{-1}$.

\section{The CUBS Program}
\label{section:data}

The CUBS program is designed to map the dominant cosmic baryon
reservoirs in intergalactic and circumgalactic space at intermediate
redshifts using QSO absorption spectroscopy, bridging the gap between
existing efforts at $z<0.4$ and at $z\approx 2$.  It utilizes the high
UV throughput and medium spectral resolution offered by the Cosmic
Origins Spectrograph (COS; Green \etal\ 2012) on board {\it HST} for
probing the physical conditions and chemical content of diffuse gas
based on observations of a suite of absorption transitions from
different ions.  Specifically, COS with the G130M and G160M gratings
provides spectral coverage over $\lambda=1100$--$1800$ \AA\ for a wide
range of ionic transitions at $z<1$.  These include the hydrogen
Lyman-series transitions and heavy elements such as carbon, nitrogen,
oxygen, silicon, sulfur, etc.\ in several ionization states (examples are 
presented in Boettcher \etal\ 2020, Cooper \etal\ 2020, Johnson \etal\ 2020, 
and Zahedy \etal\ 2020 in preparation).  An added bonus of targeting the redshift range at $z>0.4$
is the ability to precisely measure the neutral hydrogen column
density, $N({\rm HI})$, using the higher-order Lyman series absorption
transitions (e.g., Rudie \etal\ 2013; Chen \etal\ 2018; Zahedy
\etal\ 2019).

In addition, a critical component of the CUBS program is a
comprehensive deep galaxy survey in the fields around these 15 QSOs.
The locations and properties of galaxies, together with the absorption
properties of associated halos from the COS spectra, provide direct
constraints for feeding and feedback in galactic halos.  Our galaxy
survey is carried out both in space using the Wide Field Camera 3
(WFC3) and the IR channel in parallel slitless grism mode, and on the
ground using the VLT and Magellan telescopes.  The slitless grism
spectroscopy utilizes the G102 and G141 grisms to target nebular lines
such as [\OIII], H$\beta$, and H$\alpha$ at $z\apl 1$ from galaxies as
faint as $AB(H)\approx 23-24$ mag.  While the field will be offset by
3--4 Mpc from the QSO, it complements the ground-based galaxy survey for
mapping the large-scale structures in the QSO fields.

The ground-based galaxy survey consists of three components: (1) a
{\it shallow and wide} component using the IMACS multi-object imaging
spectrograph (Dressler et al.\ 2006) on the Magellan Baade telescope
to target $L_*$ galaxies at $z_{\rm gal}\apl 0.8$ and angular
distances at $\theta\apl\,10'$ (corresponding to $d<3$--5 physical
Mpc) from the QSO sightline, (2) a {\it deep and narrow} component
using the Low Dispersion Survey Spectrograph 3 (LDSS3) on the Magellan
Clay telescope to target all galaxies as faint as 0.1\,$L_*$ up to
$z$\,$\approx$\,1 (and fainter at lower redshifts) at $\apl\,3'$ in
angular radius (corresponding to $d<300$--500 physical kpc; pkpc) from
the QSO sightline, and (3) an {\it ultradeep} component using MUSE on
the VLT UT4 to target galaxies as faint as $\approx 0.01\,L_*$ at
$z$\,$\approx$\,1 at $\apl\,30''$ in angular radius (corresponding to
$d<250$ physical kpc) from the QSO sightline.  The {\it shallow and
  wide} component will enable large-scale ($\approx 1$--10 Mpc)
cross-correlation studies between gas and galaxies.  The {\it
  ultradeep} and {\it deep and narrow} components enable detailed
studies of gas flows and the chemical enrichment in the CGM at
projected distances $d\,\apl\,300$ pkpc from galaxies with mass as low
as $M_{\rm star}\,\sim\,10^9\,\msun$ at $z\apl 1$.  The primary
scientific objectives are: (1) to measure the cosmic mass density
evolution of heavy ions; (2) to determine the metallicity and
ionization state of the diffuse CGM and IGM; (3) to constrain the
origin and evolution of the chemically-enriched CGM in halos of
different masses and star formation histories; and (4) to investigate
the environmental effects in distributing heavy elements beyond galaxy
halos.  Here we describe the program design and associated
spectroscopic observations.

\begin{table}
\scriptsize
\centering
\caption{Summary of the CUBS QSO Sample}
\label{table:sample}
\centering {
\begin{tabular}{lrrcccc}
\hline \hline
\multicolumn{1}{c}{} & \multicolumn{1}{c}{} & \multicolumn{1}{c}{} & \multicolumn{1}{c}{} & \multicolumn{1}{c}{FUV} & \multicolumn{1}{c}{NUV} & \multicolumn{1}{c}{} \\
\multicolumn{1}{c}{QSO} & \multicolumn{1}{c}{RA(J2000)} & \multicolumn{1}{c}{Dec(J2000)} & \multicolumn{1}{c}{$z_{\rm QSO}$} & \multicolumn{1}{c}{(mag)} & \multicolumn{1}{c}{(mag)} & \multicolumn{1}{c}{Ref.$^a$}  \\
\hline 
J0028$-$3305 & 00:28:30.405 & $-$33:05:49.25 & 0.887 & 17.33 & 16.52 & (1) \\
J0110$-$1648 & 01:10:35.511 & $-$16:48:27.70 & 0.777 & 17.31 & 16.72 & (2) \\
J0111$-$0316 & 01:11:39.171 & $-$03:16:10.89 & 1.234 & 18.47 & 16.66 & (3) \\
J0114$-$4129 & 01:14:22.123 & $-$41:29:47.29 & 1.018 & 18.33 & 16.71 & (4) \\
J0119$-$2010 & 01:19:56.091 & $-$20:10:22.73 & 0.812 & 16.86 & 16.15 & (5) \\
J0154$-$0712 & 01:54:54.682 & $-$07:12:22.17 & 1.289 & 17.07 & 16.40 & (3) \\
J0248$-$4048 & 02:48:06.286 & $-$40:48:33.66 & 0.883 & 16.11 & 15.47 & (4) \\
J0333$-$4102 & 03:33:07.076 & $-$41:02:01.15 & 1.124 & 17.60 & 16.33 & (4) \\
J0357$-$4812 & 03:57:21.918 & $-$48:12:15.16 & 1.016 & 17.76 & 16.84 & (6) \\
J0420$-$5650 & 04:20:53.907 & $-$56:50:43.96 & 0.944 & 17.61 & 16.86 & (4) \\
J0454$-$6116 & 04:54:15.952 & $-$61:16:26.56 & 0.784 & 16.89 & 16.16 & (3) \\
J2135$-$5316 & 21:35:53.202 & $-$53:16:55.82 & 0.806 & 17.13 & 15.94 & (3) \\
J2308$-$5258 & 23:08:37.796 & $-$52:58:48.94 & 1.067 & 17.97 & 16.73 & (4) \\
J2339$-$5523 & 23:39:13.218 & $-$55:23:50.84 & 1.354 & 17.91 & 16.37 & (4) \\
J2245$-$4931 & 22:45:00.207 & $-$49:31:48.46 & 1.003 & 18.10 & 16.90 & (3) \\
\hline
\multicolumn{6}{l}{$^a$Referenes: (1) Lamontagne \etal\ (2000); (2) Perlman \etal\ (1998);}\\
\multicolumn{6}{l}{\ \ (3) Monroe \etal\ (2016); (4) Wisotzki \etal\ (2000); (5) Jones \etal\ }\\
\multicolumn{6}{l}{\ \ (2009); (6) Savage \etal\ (1978).} \\
\end{tabular}
}
\end{table}

\subsection{Program Design}

To facilitate a systematic and unbiased study of the CGM/IGM at
$z\approx 0.4$--1, the CUBS QSOs are selected to be at $z_{\rm
  QSO}\apg 0.8$ and bright in the GALEX near-UV bandpass (NUV;
1770--2730 \AA).  The QSOs are selected from existing surveys with
available spectra for redshift confirmations, including the
Hamburg/ESO survey (Wisotzki \etal\ 2000), the Sloan Digital Sky
Survey (SDSS; York \etal\ 2000; Eisenstein \etal\ 2011), the
Ultraviolet-bright Quasar Survey (UVQS; Monroe \etal\ 2016), and our
own spectroscopic observations for confirmations.  Redshift
uncertainties are typically $dz/(1+z)\approx 0.002$ for the UVQS and
significantly better for SDSS (e.g., Hewett \& Wild 2010).  By
targeting QSOs at $z_{\rm QSO}\!=\!0.8$--1.3, we optimize the survey
efficiency and maximize the survey pathlength offered by the spectral
coverage of COS for each QSO sightline.

The NUV magnitude limited QSO
selection criterion is motivated by the expectation that the presence
of a LLS or a pLLS at $z\apl 0.9$ attenuates the background QSO light
in the far-UV channel (FUV; 1350--1780 \AA; see the top panel of
Figure 1).  Consequently, targeting known FUV-bright QSOs at $z_{\rm
  QSO}\apg 0.8$ would impose a bias against sightlines intercepting a
LLS or partial LLS at lower redshifts.  Finally, we select the QSOs
from regions covered by the Dark Energy
Survey\footnote{https://www.darkenergysurvey.org/} (DES; e.g.,
Drlica-Wagner \etal\ 2018) on the ground.  The available DES $g, r, i,
z, Y$ images are supplemented with deeper $g$, $r$, and near-infrared
$H$-band images from the Magellan Telescopes to enable systematic
studies of galaxy environments of individual absorbers.


To obtain a representative map of the dominant cosmic baryon reservoir
at $z=0.4$--1, the targeted sample size is defined such that (1) a
statistically representative sample of ${\approx 100}$ of
$0.1\,L_*$--$L_*$ galaxies at ${z>0.4}$ can be established for a
comprehensive study of the CGM and (2) a large redshift pathlength is
reached for IGM metal-absorption line surveys.  Based on the best-fit
luminosity functions for red and blue galaxies from Cool et
al. (2012), we estimate that 15 QSO fields are needed for establishing
a sample of $\approx 75$ blue, star-forming and $\approx 30$ red,
evolved galaxies over a wide range of luminosity at redshifts between
$z_{\rm gal}\approx 0.4$ and $z_{\rm gal}\approx 0.8$ and projected
distances $d<\,300$ pkpc from the QSO sightlines.  In
addition, a complete galaxy survey carried out in these CUBS fields
will also double the number of $z<0.4$ galaxies with known CGM
constraints.  Furthermore, combining 15 new CUBS QSOs and available
archival sightlines is expected to lead to the largest redshift survey
pathlength of $\Delta\,z\approx 8$ (13) for high-ionization species
probed by the \OVI\,$\lambda\lambda\,1031, 1037$ and
\NeVIII\,$\lambda\lambda\,770, 780$ doublets at $z>0.4$.  These
represent new samples of \OVI\ and \NeVIII\ absorbers that are
statistically significant in size for robust measurements of the
frequency distribution function and the cosmic mass density of these
highly-ionized species.  Because the absorber samples are drawn from
random sightlines, they probe diverse gaseous environments (i.e., from
interstellar to circumgalactic and intergalactic space).

\begin{figure}
  \centering
\includegraphics[scale=0.3,angle=0]{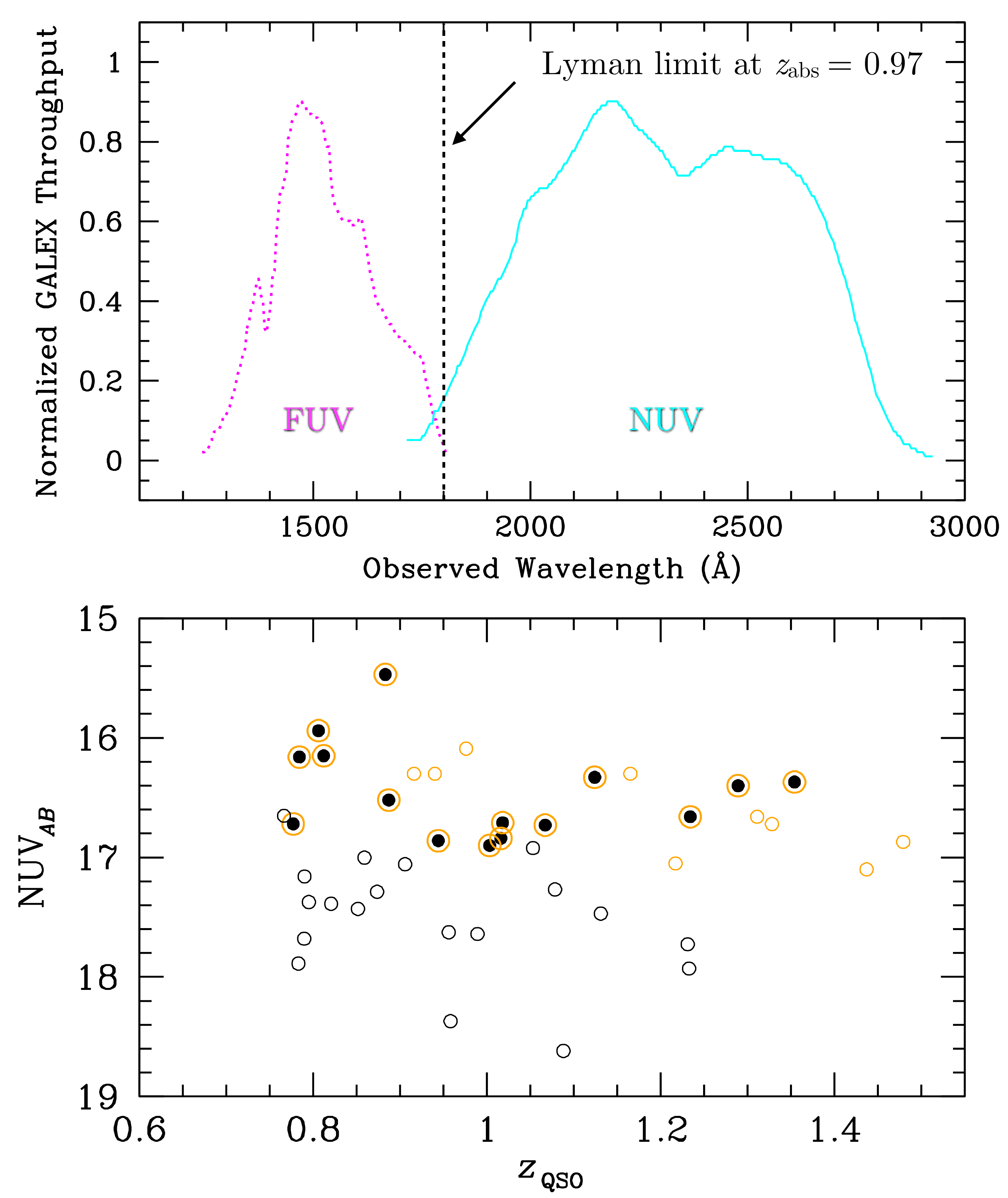}
\caption{{\it Top}: Normalized throughput functions for the GALEX FUV
  (dotted magenta curve) and NUV (solid cyan curve) bandpasses.  LLSs
  at $z_{\rm abs}\apl 0.97$ attenuate the observed FUV flux in the
  background QSO.  A FUV-bright QSO sample would therefore be biased
  against these low-redshift LLSs.  Therefore, CUBS QSOs are selected
  based on their NUV brightness.  {\it Bottom}: NUV magnitude versus
  redshift distribution of $z_{\rm QSO}>0.7$ QSOs with UV spectra
  available in the {\it HST} archive.  Those with available
  high-quality COS UV spectra of S/N\,$\apg 15$ resel$^{-1}$ are
  highlighted in orange circles, and those obtained as part of the
  CUBS program are shown in solid points.}
        \label{figure:qso}
\end{figure}

The QSO fields are selected blindly without prior knowledge of the
line-of-sight galactic environment.  A lesson learned from previous
CGM experiments is that preferentially selecting QSO fields with a
larger number of known photometrically-selected galaxies does not
necessarily help increase the galaxy sample size for a fixed number of
QSO fields but likely biases the galaxy sample toward galaxy groups
(see for example Werk \etal\ 2012; Qu \& Bregman 2018).  The inclusion
of galaxy groups not only skews the galaxy sample toward more massive
halos but also introduces ambiguities in interpreting the physical
connections between the absorber and multiple group members.  The QSOs
in the CUBS program are presented in Table 1.  Figure 1 (bottom panel)
presents the NUV magnitude versus redshift distribution of known UV
bright QSOs at $z_{\rm QSO}>0.7$.  The CUBS QSOs are highlighted in
solid points, showing a three-fold increase in the number of
high-quality UV absorption spectra at $z_{\rm QSO}>0.7$.

\subsection{COS UV Spectroscopy}


Medium-resolution, high signal-to-noise ratio ($S/N$) FUV spectra of
15 new NUV bright QSOs were obtained under the CUBS program
(PID$\,=\,$15163; PI: Chen) using COS on board the {\it HST}.  COS
with the G130M and G160M gratings and a combination of multiple
central wavelength settings (see Table 2) offers a contiguous spectral
coverage of $\lambda=1100$--$1800$ \AA, with a spectral resolution of
Full-Width-at-Half-Maximum $\delta\,v_{\rm FWHM}\approx 20$ \kms\ for
observing a wide range of ionic transitions at $z<1$.  These include
the hydrogen Lyman-series transitions from \lya\ and \lyb\ onward to
the Lyman-limit transition, and heavy-element transitions such as
\OIII\,$\lambda\,702$, \NeVIII\,$\lambda\lambda\,770,780$,
\OIV\,$\lambda\,787$, \OII\,$\lambda\,834$,
\CII\,$\lambda\lambda\,903a,b$, \CIII\,$\lambda\,977$,
\OI\,$\lambda\,988$, \OVI\,$\lambda\lambda\,1031,1037$, etc.  The full
coverage of the \HI\ Lyman series enables precise and accurate
measurements of the neutral hydrogen column density $N({\rm HI})$.
The relative abundances between different ions enable accurate
estimates of the ionization state and metallicity of the gas.

All COS target acquisitions (TA) were performed using $S/N > 50$
ACQ/IMAGEs.  Analysis of the primary and confirmation images reveal
that all targets were centered to better than $0.015''$ ($0.016''$)
along the dispersion (cross-dispersion) direction.  For G130M and
G160M spectra, the dispersion velocity offsets due to TA are less than
1.5 \kms.  A summary of the COS observations is presented in Table 2,
which lists for each QSO, the name and redshift of the QSO, the total
exposure time per grating in seconds, and the mean $S/N$ per
resolution element, $\langle\,S/N\,\rangle_{\rm resel}$ in the final
combined spectrum.

\begin{table}
\scriptsize
\centering
\caption{Journal of CUBS {\it HST} COS Observations}
\label{table:cosobs}
\centering {
\begin{tabular}{lrrrc}
\hline \hline
\multicolumn{1}{c}{} & \multicolumn{1}{c}{} & \multicolumn{2}{c}{$t_{\rm exp}$ (sec)} & \multicolumn{1}{c}{} \\
\cline{3-4} \\
\multicolumn{1}{c}{QSO} & \multicolumn{1}{c}{$z_{\rm QSO}$} & \multicolumn{1}{c}{G130M$^a$} & \multicolumn{1}{c}{G160M$^b$} & \multicolumn{1}{c}{$\langle\,S/N\,\rangle_{\rm resel}$} \\
\hline 
J0028$-$3305 & 0.887 & 13253 & 17589 & 23 \\
J0110$-$1648 & 0.777 & 15320 & 19802 & 31 \\
J0111$-$0316 & 1.234 & 15149 & 19711 & 20 \\
J0114$-$4129 & 1.018 & 15320 & 19798 & 12 \\
J0119$-$2010 & 0.812 &  7049 & 13437 & 24 \\
J0154$-$0712 & 1.289 &  9716 & 14843 & 28 \\
J0248$-$4048 & 0.883 &  4516 &  5538 & 20 \\
J0333$-$4102 & 1.124 & 12534 & 14849 & 24 \\
J0357$-$4812 & 1.016 & 32664 & 24630 & 27 \\
J0420$-$5650 & 0.944 & 17797 & 20570 & 22 \\
J0454$-$6116 & 0.784 &  8330 & 13503 & 22 \\
J2135$-$5316 & 0.806 &  7098 &  9738 & 18 \\
J2245$-$4931 & 1.003 &  9395 & 19321 & 14 \\
J2308$-$5258 & 1.067 & 20618 & 22118 & 23 \\
J2339$-$5523 & 1.354 &  9388 & 14987 & 22 \\
\hline
\multicolumn{5}{l}{$^\mathrm{a}$Two central wavelength settings, C1291 and C1223, were used for a.}\\
\multicolumn{5}{l}{\ \ contiguous spectral coverage.}\\
\multicolumn{5}{l}{$^\mathrm{b}$Four central wavelength settings, C1577, C1589, C1611, and C1623}\\
\multicolumn{5}{l}{\ \ were used.}\\
\end{tabular}
}
\end{table}

Raw data from COS were reduced following standard pipeline procedures
using {\it CALCOS} (v3.3.4 or v3.3.5).  In addition, individual one-dimensional
spectra were further processed and combined using custom software to
ensure the accuracy of wavelength calibration and to optimize the
$S/N$ in the final combined spectra.  A detailed description of the
software can be found in Chen \etal\ (2018).  In summary, {\it
  relative} wavelength offsets between individual spectra were first
determined using a low-order polynomial that best describes the
offsets between common absorption lines found in different exposures.
These wavelength-corrected spectra were then coadded to form a final
combined spectrum using individual exposure times as weights.  The
final {\it absolute} wavelength calibration was guided by either the
line-of-sight velocity offset of the Milky Way \CaII\ H\&K lines or
the redshift of a strong intervening \MgII\ absorber detected in the
ground-based MIKE optical echelle spectrum of the QSO (see \S\ 2.3
below), which sets the wavelengths for the associated FUV transitions
in the COS spectra.  A mean wavelength zero point offset was then
determined by registering the associated low-ionization lines observed
in the COS spectra to the expected wavelength in vacuum.  The final
wavelength solution generated from our custom software was found to be
accurate to within $\pm 5$ \kms, based on comparisons of the velocity
centroids between low-ionization lines observed in COS spectra and
those of \MgII\,$\lambda\lambda\,2796, 2803$ lines observed in
higher-resolution ground-based echelle spectra ($\delta\,v_{\rm
  FWHM}\approx 8$--10 \kms).  Finally, each combined spectrum was
continuum-normalized using a low-order polynomial fit to spectral
regions free of strong absorption features.  The final
continuum-normalized spectra have a median $\langle\,S/N\rangle_{\rm
  resel} \approx 12$--31 (see Table 2).  The large variation in the
$S/N$ of the final combined spectra is largely due to QSO variability.
For example, comparing $S/N$ in the final COS spectra and known GALEX
magnitudes of the QSOs, we estimate that J0110$-$1648 had brightened
by a factor of $\approx 1.4$, while J0114$-$4129 had faded by a factor
of 2 since the GALEX observations.  Large QSO variability was also
directly observed in the COS spectra of J2308$-$5258 obtained several
months apart.

\subsection{Optical Echelle Spectroscopy}

\begin{table}
\scriptsize
\centering
\caption{Journal of CUBS Magellan MIKE Observations}
\label{table:mikeobs}
\centering {
\begin{tabular}{lrrrrrr}
\hline \hline
& & \multicolumn{1}{c}{$V$} &  \multicolumn{1}{c}{$t_{\rm exp}$} & \multicolumn{1}{c}{${\rm FWHM}$} & \multicolumn{2}{c}{$\langle\,S/N\,\rangle_{\rm resel}$} \\
 \cline{6-7} \\
\multicolumn{1}{c}{QSO} & \multicolumn{1}{c}{$z_{\rm QSO}$} & \multicolumn{1}{c}{(mag)} &  \multicolumn{1}{c}{(sec)} & \multicolumn{1}{c}{(km/s)} & \multicolumn{1}{c}{$3500$ \AA} & \multicolumn{1}{c}{$4500$ \AA} \\
\hline 
J0028$-$3305 & 0.887 & 16.4 & 2100 & 8 &  6 & 35 \\
J0110$-$1648 & 0.777 & 16.0 & 1800 & 8 & 12 & 38 \\
J0111$-$0316 & 1.234 & 15.5 & 1800 & 8 & 18 & 55 \\
J0114$-$4129 & 1.018 & 16.7 & 3600 & 8 & 10 & 42 \\
J0119$-$2010 & 0.812 & 15.8 & 1800 & 8 & 16 & 55 \\
J0154$-$0712 & 1.289 & 15.8 & 1500 & 8 & 14 & 42 \\
J0248$-$4048 & 0.883 & 15.1 &  900 & 8 & 13 & 48 \\
J0333$-$4102 & 1.124 & 15.8 & 2700 & 10 & 20 & 65 \\
J0357$-$4812 & 1.016 & 16.0 & 5400 & 10 &  12 & 41 \\
J0420$-$5650 & 0.944 & 16.1 & 3000 & 10 &  8 & 30 \\
J0454$-$6116 & 0.784 & 15.8 & 1800 & 10 & 17 & 58 \\
J2135$-$5316 & 0.806 & 15.8 & 3200 & 8 & 14 & 64 \\
J2245$-$4931 & 1.003 & 16.5 & 3600 & 8 &  6 & 35 \\
J2308$-$5258 & 1.067 & 16.2 & 5800 & 8 & 11 & 52 \\
J2339$-$5523 & 1.354 & 15.5 & 2700 & 8 & 15 & 78 \\
\hline
\end{tabular}
}
\end{table}

We complement the FUV spectra from COS with optical echelle spectra of
the QSOs, obtained using MIKE (Bernstein \etal\ 2003) on the Magellan
Clay telescope.  MIKE delivers an unbinned pixel resolution of
$0.12''$ ($0.13''$) along the spatial direction and $\approx 0.02$
(0.05) \AA\ along the spectral direction in the blue (red) arm,
covering a wavelength range of $\lambda=3200$--5000 (4900--9200) \AA.
It provides extended spectral coverage for additional ionic transitions
through observations of the \FeII\ absorption series, the
\MgII\,$\lambda\lambda\,2796, 2803$ doublet features,
\MgI\,$\lambda\,2852$, and \CaII\,$\lambda\lambda\,3934, 3969$
absorption, and enables accurate relative abundances studies (e.g.,
Zahedy \etal\ 2016, 2017).

The majority of the optical echelle spectra of the CUBS QSOs were
obtained between September 2017 and March 2018, with additional
observations taken in February 2019 and October 2019.  These UV-bright
QSOs are also bright in the optical window, with $V$-band magnitude
ranging from $V=15.1$ mag to $V=16.5$ mag.  The echelle spectroscopy of
CUBS QSOs was carried out as a filler program within other regular
programs.  As a result, two readout settings, $2\times 2$ versus
$3\times 3$ binning, were adopted for the QSO sample, leading to a
spectral resolution of $\delta\,v_{\rm FWHM}\approx 8$ \kms, and 10 \kms,
respectively.  The echelle spectra were processed and extracted using
custom software described in Chen \etal\ (2014) and in Zahedy
\etal\ (2016).  Wavelength calibrations were performed using a ThAr
frame obtained immediately after each science exposure and
subsequently corrected to a vacuum and heliocentric wavelength scale.
Relative flux calibrations were performed using a sensitivity function
determined from a spectrophotometric standard star observed on the
same night as the CUBS QSOs.  Individual flux-calibrated echelle
orders from different exposures were then coadded and combined to form
a single final spectrum.  Finally, the combined spectrum was
continuum-normalized using a low-order polynomial fit to the spectral
regions free of strong absorption features.

A summary of available optical echelle spectra is presented in Table
3, which lists for each QSO the $V$-band magnitude, the total
accumulated exposure time, spectral resolution, and the mean $S/N$ per
resolution element at $\lambda =3500$ and 4500 \AA.  The mean
$S/N_{\rm resel}$ of the final combined spectra ranges between 6 and
20 at $\lambda=3500$ \AA\ and between 30 and $\approx 80$ at
$\lambda=4500$ \AA.  These echelle spectra offer a factor of two
larger resolving power for metal lines than the FUV spectra from
COS.  They serve as an important guide for analyzing the COS spectra.

\subsection{MUSE Observations}

\begin{table}
\scriptsize
\centering
\caption{Journal of Completed MUSE-LLS Observations}
\label{table:mikeobs}
\centering {
\begin{tabular}{lrrrr}
\hline \hline
 & \multicolumn{1}{c}{$t_{\rm exp}$} & \multicolumn{1}{c}{FWHM} & \multicolumn{1}{c}{$AB(r)^a$} & \multicolumn{1}{c}{SB(7000 \AA)$^b$} \\
\multicolumn{1}{c}{QSO} &  \multicolumn{1}{c}{(sec)} &  \multicolumn{1}{c}{(arcsec)} & \multicolumn{1}{c}{(mag)} & \multicolumn{1}{c}{${\rm erg}/{\rm s}/{\rm cm}^{2}/{\rm \AA}/{\rm arcsec}^{2}$} \\
\hline 
J0248$-$4048$^c$ & 7650 & 0.7 & 27.0 &  $9.8\times 10^{-20}$  \\
J0357$-$4812 & 9390 & 0.6 & 27.4 & $7.5\times 10^{-20}$ \\
J2135$-$5316 & 6840 & 0.6 & 26.9 & $1.3\times 10^{-19}$ \\
\hline
\multicolumn{5}{l}{$^\mathrm{a}$5-$\sigma$ limiting magnitude in the pseudo $r$-band integrated from $6000$ \AA}\\
\multicolumn{5}{l}{\ to $7000$ \AA.} \\
\multicolumn{5}{l}{$^\mathrm{b}$1-$\sigma$ limiting surface brightness  at 7000 \AA\ per sq.\ 
arcsecond aperture}\\
\multicolumn{5}{l}{$^\mathrm{c}$One of the OBs for J0248$-$4048 was obtained through clouds.  While}\\
\multicolumn{5}{l}{\ the exposures through clouds do not reach the same depth as those}\\
\multicolumn{5}{l}{\ obtained under clear skies, including all exposures yields the deepest}\\
\multicolumn{5}{l}{\ combined data cube for this field.} \\
\end{tabular}
}
\end{table}

\begin{figure*}
\includegraphics[scale=0.45]{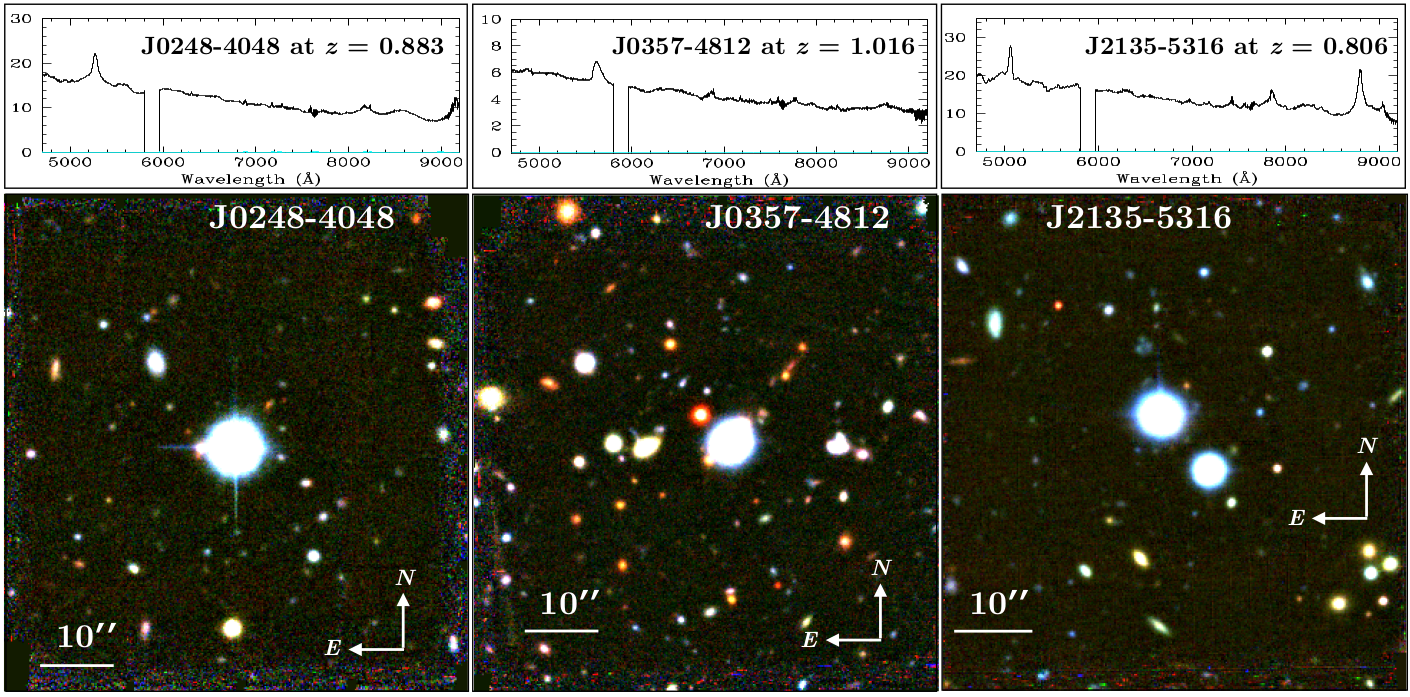}
\caption{MUSE observations of three CUBS QSO fields.  Optical spectra
  of the QSOs extracted from the combined MUSE data cube are presented
  at the {\it top}, showing the spectral coverage of MUSE.  The
  spectral gap at $5800$--5965 \AA\ in the top panels is due to the
  sodium filter, which was applied to block the scattered light from
  the laser beams for wavefront corrections.  Continuum images of the
  QSO fields are presented at the {\it bottom}.  The color images are
  produced from integrating the MUSE cube over wavelength ranges of
  $4800$--5800 \AA\ (pseudo $g$-band), $6000$--7000 \AA\ (pseudo
  $r$-band), and $7500$--8500 \AA\ (pseudo $i$-band).  For each field,
  north is up and east is to the left.  The QSO is at the center, and
  the horizontal bar in the lower-left corner marks $10''$ on the
  sky. }
\label{figure:muse}
\end{figure*}

The {\it ultradeep} galaxy survey component described at the beginning of \S\ 2 is being carried out using the Multi-Unit Spectroscopic Explorer (MUSE; Bacon et
al.\ 2010) on the VLT UT4
in service mode
under program ID, 0104.A-0147 (PI: Chen).  MUSE observes a field of
$1'\times 1'$ with a plate scale of $0.2''$ and 1.25 \AA\ per pixel,
covering a spectral range from 4800 \AA\ to 9200 \AA\ with a spectral
resolution of $\delta\,v_{\rm FWHM}\approx 120$ \kms\ at 7000 \AA.
The combined spatial and spectral resolving power of MUSE provides
high sensitivity and high efficiency for surveys of distant faint
galaxies and line-emitting nebulae, and is uniquely suitable for
uncovering faint emission close to QSO sightlines (e.g., Schroetter \etal\ 2016; Bielby
\etal\ 2017; Johnson \etal\ 2018; P\'eroux \etal\ 2019; Chen
\etal\ 2019a,b).  For the CUBS QSO fields, the combination of {\it
  HST} COS absorption spectroscopy and MUSE integral-field
spectroscopy provides a powerful tool for studying the complex
interplay between gas and galaxies at $z\apl 1$.

The CUBS-MUSE program aims to uncover galaxies as faint as $\approx
0.01\,L_*$ at $z\apl 1$ near the QSO sightlines.  The observations are
carried out in wide-field mode (WFM) with adaptive optics assistance.
While the use of lasers for wavefront corrections imposes a gap in the
spectral coverage from 5800 to 5965 \AA, it ensures a uniform mean
image quality of ${\rm FWHM}<0.8''$ across all fields.  For each CUBS
QSO field, the MUSE observations are carried out in a series of two to
three observing blocks (OBs) with individual OBs consisting of three
exposures of 850 to 1130 s each.  To optimize cosmic-ray rejection and
minimize a fixed residual flat-field pattern at the edges of
individual slicers, a small dither of $\approx 2''$--$4''$ and a
relative field rotation of 90$^\circ$ are applied to every successive
exposure.  While the program is ongoing, here we present results from
three completed CUBS QSO fields, all with a LLS found in the
foreground (see \S\ 3 below).  The QSO fields and the total
accumulated exposure time with MUSE are presented in the first two
columns of Table 4.

All MUSE data cubes are reduced using a combination of the standard
ESO MUSE pipeline (Weilbacher \etal\ 2014) and CUBEXTRACTOR, a custom
package developed by S.\ Cantalupo (see Cantalupo \etal\ 2019 for a
detailed description).  For each OB, both raw science exposures and
the associated raw calibration files, including bias, flats,
comparison arc files, and spectrophotometric standard, are retrieved
from the ESO science archive.  The ESO MUSE pipeline first generates a
master bias and a master flat, processes the arcs for wavelength
calibration and for measuring the instrument line spread function, and
twilight flats for illumination corrections.  Then it applies these
calibrations to the standard star and science frames to produce a
response function, telluric corrections, and a sky continuum using the
20\% darkest pixels in the field of view.  Finally, the pipeline
produces a sky-subtracted 3D data cube for each raw science exposure,
along with associated pixel tables storing all calibration parameters,
and a whitelight image for object identifications.  Individual data
cubes are then registered using common objects identified in the
corresponding whitelight image to form a final combined cube for each
OB.

The final product from the standard ESO pipeline exhibits apparent
residual patterns in the sky background due to imperfect illumination
correction and sky subtraction (see Figure 1 of Lofthouse et al.\ 2020
for an example).  The residuals are most severe near/at bright sky
lines, making robust identifications of faint features challenging.
In particular, MUSE consists of 24 integral field units (IFUs) and
each IFU consists of four stacks of 12 slices.  Any temporal and
wavelength-dependent changes in the illumination pattern and
instrument line spread function result in large background residuals
in the final data cube.  To improve upon the standard pipeline
reduction, the CUBEXTRACTOR package first adopts the pipeline produced
data cube with sky subtraction turned off to define a reference grid
for the post-processing steps and resamples the pixel tables on to the
reference grid.  Then ``CubeFix'' works in the reconstructed data cube
space and empirically determines a wavelength-dependent illumination
pattern per image slice using the sky background observed in each
science frame.  Next, ``CubeSharp'' performs a flux-conserving (across
the field of view) sky subtraction routine using an empirical line
spread function constructed from the skylines recorded in the science
frame.  These steps are performed twice using an updated object
catalog for masking continuum sources in the second iteration of sky
removal.  Finally, individually corrected data cubes from all OBs are
combined to form a final data cube.  Because the MUSE wavelength
solution is calibrated in air and QSO absorption spectra are
calibrated to vacuum, the product of CUBEXTRACTOR is further resampled
to vacuum wavelength to facilitate an accurate redshift comparison
between galaxies identified in MUSE and absorbers identified in the
COS spectra.


The QSO spectra and RGB images constructed using pseudo $g$-, $r$-,
and $i$-band images integrated over wavelength windows of 4800--5800
\AA, 6000-7000 \AA, and 7500--8500 \AA, respectively, from the final
MUSE data cubes of the three CUBS QSO fields are presented in Figure
2.  The image quality in the final combined cubes ranges between ${\rm
  FWHM}\approx 0.6''$ and $0.7''$ at 7000 \AA.  The image quality is
slightly worse at shorter wavelengths, ranging from $\approx 0.7''$ to
$0.8''$ at 5000 \AA, and better at longer wavelengths, ranging from
$\approx 0.56''$ to $0.6''$ at 9000 \AA.  The 5-$\sigma$ limiting
magnitude in the pseudo $r$-band ranges between $r=27.0$ and 27.4 mag
over $1''$ diameter aperture and the 1-$\sigma$ limiting surface
brightness reaches $(0.8-1.3)\times 10^{-19}\,{\rm erg}\,{\rm
  s}^{-1}\,{\rm cm}^{-2}\,{\rm \AA}^{-1}\,{\rm arcsec}^{-2}$ at 7000
\AA\ per square arcsecond aperture (see Table 4).

\section{Survey of Lyman limit absorbers at $z<1$}

The CUBS QSOs are selected based on the QSO emission redshift with
$z_{\rm QSO}\apg 0.8$ and the NUV brightness with ${\rm NUV}\apl 16.9$
(Figure 1), with no prior knowledge of the line-of-sight absorption
properties or galactic environment.  Therefore, the CUBS QSO sample
provides a uniform sample (cf.\ archival samples from Ribaudo
\etal\ 2011 or Shull \etal\ 2017) for studying the incidence of
optically-thick gas at $z\apl 1$ as well as the connection between LLS
and galaxy properties.  The COS spectra described in \S\ 2.2 are of
sufficiently high quality with $S/N_{\rm resel}>12$ to enable a robust
identification of both the Lyman discontinuity at $\approx 912$
\AA\ and the Lyman series lines at longer wavelengths.  Here we
describe the procedures we use to identify these absorbers and to
measure $N(\HI)$.

\subsection{The search for Lyman continuum breaks}

The search for LLS in the CUBS QSO sample is carried out using the
following steps.  For each QSO sightline, a minimum redshift $z_{\rm
  min}$ for the LLS search is defined by the minimum wavelength where
$S/N_{\rm resel}>3$, while the maximum survey redshift $z_{\rm max}$
is defined by either the maximum wavelength of the COS spectrum or the
emission redshift of the QSO, excluding the velocity window of
$|\Delta\,v|=3000$ \kms\ from $z_{\rm QSO}$ to avoid the QSO proximity
zone where the ionizing radiation intensity is expected to be enhanced
due to the background QSO (e.g., Pascarelle \etal\ 2001; Wild
\etal\ 2008).  Table 5 summarizes $z_{\rm min}$ and $z_{\rm max}$ for
each CUBS QSO, along with the redshift survey pathlength
$\Delta\,z_{\rm LL}$, in columns (2)--(4).  Together, the 15 CUBS QSOs
provide a total redshift survey pathlength of $\Delta\,z_{LL}=9.3$ for
new LLSs.

\begin{table}
\scriptsize
\centering
\caption{Summary of new (p)LLSs in the CUBS fields}
\label{table:sample}
\centering {
\begin{tabular}{lrrrcr}
\hline \hline
\multicolumn{1}{c}{Field} & \multicolumn{1}{c}{$z_{\rm min}$} & \multicolumn{1}{c}{$z_{\rm max}$} & \multicolumn{1}{c}{$\Delta\,z_{\rm LL}$} & \multicolumn{1}{c}{$z_{\rm abs}$} & \multicolumn{1}{c}{$\tau_{\rm 912}$} \\
\multicolumn{1}{c}{(1)} & \multicolumn{1}{c}{(2)} & \multicolumn{1}{c}{(3)} & \multicolumn{1}{c}{(4)} & \multicolumn{1}{c}{(5)} & \multicolumn{1}{c}{(6)} \\
\hline
J0028$-$3305 & 0.21 & 0.87 & 0.66 &    ...    &       ...        \\
J0110$-$1648 & 0.20 & 0.76 & 0.56 & 0.4723$^a$ & 0.18$\pm$0.01 \\
             &      &      &      & 0.5413$^a$ & 0.21$\pm$0.01 \\
J0111$-$0316 & 0.57 & 0.94 & 0.37 & 0.5762$^b$ & $>6.61$ \\
J0114$-$4129 & 0.23 & 0.94 & 0.71 & 0.3677 & 0.16$\pm$0.02 \\
             &      &      &      & 0.9001 & 0.41$\pm$0.03 \\
J0119$-$2010 & 0.21 & 0.79 & 0.58 &   ...     &       ...       \\
J0154$-$0712 & 0.20 & 0.94 & 0.74 & 0.3743 & 0.24$\pm$0.01 \\
J0248$-$4048 & 0.24 & 0.86 & 0.62 & 0.3640 & 2.48$\pm$0.01 \\
J0333$-$4102 & 0.20 & 0.94 & 0.74 & 0.9372 & 0.61$\pm$0.02 \\
J0357$-$4812 & 0.21 & 0.94 & 0.73 & 0.4353 & 0.99$\pm$0.01 \\
J0420$-$5650 & 0.21 & 0.92 & 0.71 &   ...     &       ...        \\
J0454$-$6116 & 0.21 & 0.76 & 0.55 &   ...     &       ...        \\
J2135$-$5316 & 0.62 & 0.79 & 0.17 & 0.6226 & $>6.27$ \\
J2245$-$4931 & 0.22 & 0.94 & 0.72 &   ...     &       ...        \\
J2308$-$5258 & 0.22 & 0.94 & 0.72 & 0.2603 & 0.58$\pm$0.02 \\
             &      &      &      & 0.5427 & 2.53$\pm$0.01 \\
J2339$-$5523 & 0.21 & 0.94 & 0.73 &   ...     &       ...        \\
\hline
\multicolumn{6}{l}{$^a$see Cooper \etal\ (2020)} \\
\multicolumn{6}{l}{$^b$see Boettcher \etal\ (2020)} \\
\end{tabular}
}
\end{table}

\begin{figure*}
\includegraphics[scale=0.4,angle=0]{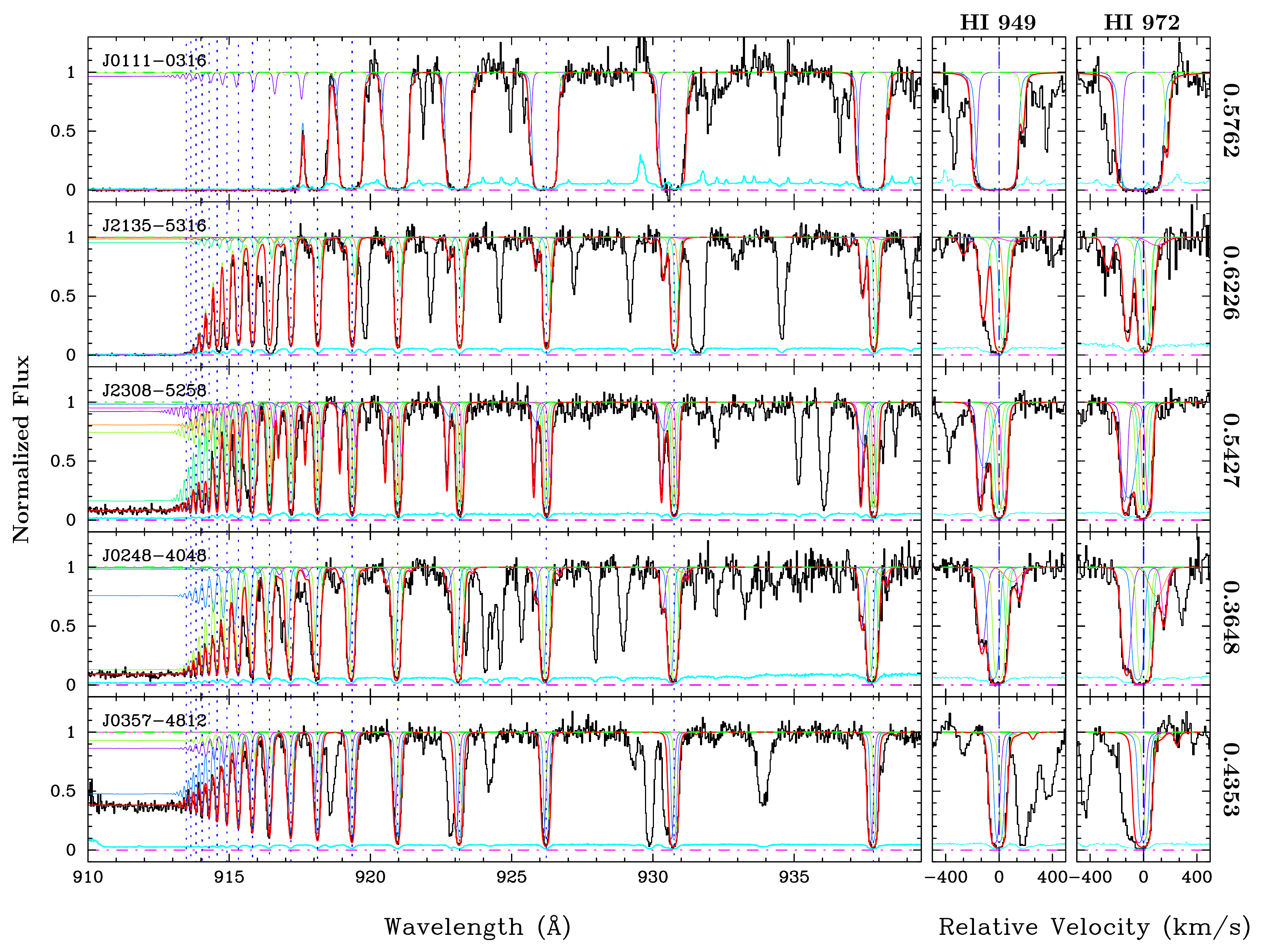}
\caption{The Lyman series absorption spectra of the five new
  optically-thick \HI\ absorbers with $\tau_{\rm 912}\apg 1$ found in
  the CUBS QSO sightlines.  The absorbers are ordered with decreasing
  \lnhone\ from top to bottom.  The continuum-normalized spectra are
  shown in black with the corresponding 1-$\sigma$ error shown in
  cyan.  The green and magenta dash-dotted lines mark the normalized
  continuum and zero flux levels for guidance.  For each absorber, the
  velocity profiles of Ly$\delta$ (\HI\,949) and Ly$\gamma$ (\HI\,972)
  are presented in the two right panels with zero velocity
  corresponding to the strongest \HI\ component in Table 6.  The
  remaining higher-order Lyman series lines, along with the Lyman
  limit, are presented in the left panel with the vertical blue dotted lines
  indicating the expected positions of the Lyman transitions.  The
  best-fit Lyman series spectra are shown in red with individual
  components displayed in different colors.}
\label{figure:LLS}
\end{figure*}

Each Lyman limit absorber is then identified based on an apparent flux
discontinuity in the QSO spectrum and verified based on the presence
of associated Lyman series lines.  A total of 12 such absorbers are
found along nine of the 15 QSO sightlines, with the remaining six
sightlines displaying no evident continuum breaks in the COS spectral
window\footnote{We note the presence of a likely LLS at $z_{\rm
    abs}\approx 1$ toward J2339$-$5523 based on the observed spectral
  slope of the QSO and a suite of absorption transitions in the COS
  data.  This system will be presented in Johnson et al.\ (in
  preparation).}.  A mean opacity, $\tau_{912}$, is determined based
on the observed flux decrement at 911.76 \AA\ relative to the expected
continuum flux from extrapolating a linear model that best describes
the continuum at rest-frame 920--923 \AA\ at the redshift of the
absorber, $z_{\rm abs}$.  The error in $\tau_{912}$ is estimated
including uncertainties in both the continuum model and the
measurement uncertainties in the mean flux observed at rest-frame
911.76 \AA.  The results, including $z_{\rm abs}$, are presented in
columns (5)--(6).  Of the 12 absorbers identified based on an apparent
Lyman discontinuity, five are LLSs with $\tau_{912}\apg 1$ at $z_{\rm
  abs}=0.36$--0.62 and five are partial LLSs (pLLSs) with $\tau_{912}$
from 0.2 to $\apl 1$ at $z_{\rm abs}=0.26$--0.94.  Figure 3 presents
the full Lyman series spectra of the five new LLSs from this search in
descending order of $N(\HI)$.

\subsection{Measurements of $N(\HI)$ and $b_{\scriptsize {\rm HI}}$}

In addition to the prominent Lyman discontinuity, Figure 3 also shows
that each of these new LLSs is resolved into multiple components of
varying absorption strength.  The resolved component structure is
clearly displayed in the velocity profiles of both \HI\ and the
associated metal lines presented in Figure 4.  In particular,
higher-resolution ($\delta\,v_{\rm FWHM}\approx 8$ \kms) ground-based
optical echelle spectra show that the associated \MgII\ doublets of
these LLSs are resolved into between two and six well-defined
components.

To obtain accurate measurements of $N(\HI)$ for individual components,
we perform a Voigt profile analysis that takes into account the full
Lyman series lines and the observed flux discontinuity at the Lyman
limit (see also Chen \etal\ 2018; Zahedy \etal\ 2019).  We first
generate a model absorption spectrum based on the minimum number of
discrete components required to explain the observed absorption
profiles.  This process is guided by the component structure of the
associated \MgII\ doublet for each LLS.  Each component is
characterized by three parameters: $N(\HI)$, $b_{\rm HI}$, and the
velocity centroid $dv_c$ relative to the redshift of the strongest
\HI\ absorbing component $z_{\rm abs}$.  The model Lyman series
spectrum is then convolved with the COS line spread function (LSF)
appropriate for Lifetime Position 4, at which the spectra were
recorded.  Next, we perform a $\chi^2$ minimization routine to
determine the best-fit model parameters by comparing the LSF-convolved
model spectrum with observations.  For \HI\ components with associated
\MgII, $dv_c$ is fixed at the location determined from the centroid of
the \MgII\ component.  For \HI\ components without detected \MgII,
$dv_c$ is allowed to vary during the $\chi^2$ minimization
routine\footnote{One exception is the absorber at $z_{\rm abs}=0.5762$
  toward J0111$-$0316, which turns out to be an ${\rm H}_2$-bearing
  DLA, for which all available Lyman series lines are highly
  saturated, preventing us from resolving individual \HI\ components
  within $|d\,v_c|\apl 150$ \kms, while the associated metal lines,
  such as the \MgII\ doublet, are resolved into seven discrete
  components in available optical echelle spectra (see Boettcher
  \etal\ 2020) for details. }.

\begin{figure*}
  \includegraphics[scale=0.45]{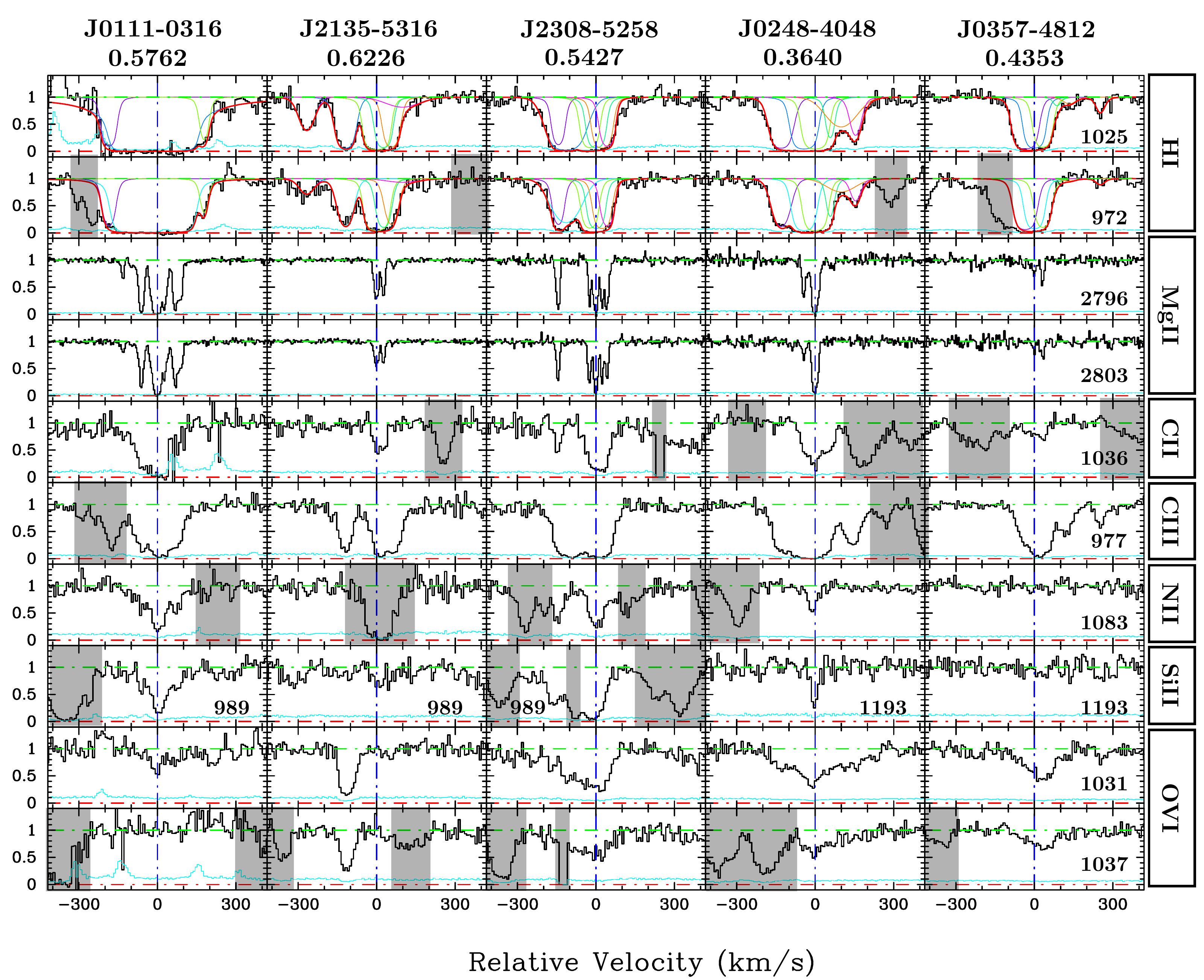}
\caption{Absorption profiles of \lyb\ and Ly$\gamma$, and associated
  \MgII, \CII, \CIII, \NII, \SiII, and \OVI\ found in the
  LLSs presented in Figure 3. The absorbers are ordered with
  decreasing \lnhone\ from left to right.  The rest-frame wavelength
  of each transition is listed in the lower-right corner of each panel
  in the right column.  Following Figure 3, zero velocity
  corresponding to the redshift of the strongest \HI\ component in
  Table 6, 1-$\sigma$ errors are shown in cyan, and spectral regions
  that are contaminated with blended absorption features are greyed
  out for clarity.  In particular, the position where the
  \NII\,$\lambda\,1083$ line is expected for the LLS at $z_{\rm
    abs}=0.6226$ toward J2135$-$5316 is dominated by a strong
  \lya\ absorber at $z_{\rm abs}=0.4468$.  The best-fit Voigt profiles
  of \lyb\ and Ly$\gamma$ lines, both separately for individual
  components (thin lines) and together for all components combined
  (thick red line), of each absorber are also reproduced in the two
  first rows for direct comparisons with resolved metal-line
  components.  }
\label{figure:ions}
\end{figure*}

To estimate the uncertainties associated with the best-fit parameters,
we perform a Markov Chain Monte Carlo (MCMC) analysis using the
\textsc{emcee} package (Foreman-Mackey \etal\ 2013).  The MCMC
analysis consists of 300 steps with an ensemble of 250 walkers,
initialized over a small region in the parameter space around the
minimum $\chi^2$ value.  The first 100 steps of each walker are
discarded when constructing a probability distribution function for
each best-fit model parameter from combining results from all 250
walkers.  The MCMC approach enables a robust evaluation of correlated
errors between blended components over a reasonable amount of
computing time.  The results of the Voigt profile analysis are
summarized in Table 6, where for each LLS the best-fit redshift of the
strongest \HI\ component $z_{\rm abs}$ and the total $N(\HI)$ summed
over all components are listed, along with the best-fit velocity
centroid $dv_c$ relative to $z_{\rm abs}$, $N_c(\HI)$, $b_c(\HI)$ and
associated 1-$\sigma$ uncertainties for individual components.

The best-fit Voigt profiles of individual \HI\ components are shown in
Figure 3 for the Lyman series from Ly$\gamma$ to the Lyman break, with
the red spectrum representing the integrated profile over all
components.  The best-fit models are also displayed in the top two
rows of Figure 4 for Ly$\beta$ and Ly$\gamma$ to contrast the velocity
structures displayed in the associated ionic transitions.  In all five
LLSs, a dominant component, containing between 75\% and 98\% of the
total \HI\ column density, is needed to explain the high-order Lyman
series profiles, while additional satellite components are needed to
explain the widths and asymmetric profiles observed in lower-order
Lyman lines such as \lya\ through Ly$\delta$.  With the exception of
the LLS at $z_{\rm abs}=0.4353$ toward J0357$-$ 4812, the dominant
\HI\ component also corresponds to the strongest component observed in
low-ionization transitions such as \CII, \NII, \MgII, and \SiII.  In
particular, the associated \MgII\ components are fully resolved in the
ground-based optical echelle spectra.  To facilitate direct
comparisons between $N(\HI)$ and ion abundances of different
components, we also present initial measurements of \MgII\ component
column densities, $N_c(\MgII)$, in Table 6, but details regarding the
Voigt profile analysis of metal absorption lines are presented in
Zahedy \etal\ (2020, in preparation).  The best-fit $b_c(\HI)$ of the
dominant \HI\ component ranges between $b_c(\HI)\approx 14$ \kms\ and
$b_c(\HI)\approx 20$ \kms, with the exception of the ${\rm
  H}_2$-bearing DLA at $z_{\rm abs}=0.5762$ toward J0111$-$0316 for
which the \HI\ components are fully blended (see Boettcher
\etal\ 2020).  Because $b_c(\HI)\equiv \sqrt{12.9^2\,T_4+b_{\rm
    turb}^2}$ \kms\ for \HI\ gas of temperature $T_{\rm gas}\equiv
10^4\times T_4$ and turbulent width $b_{\rm turb}$, the best-fit
values constrain the underlying gas turbulence or bulk motion to be
$b_{\rm turb}< 15$ \kms\ for optically-thick absorbers of $T_{\rm
  gas}\sim 10^4$ K and still smaller for warmer temperatures.

\section{Descriptions of Individual Systems}

The LLS survey described in \S\ 3 has yielded five new LLSs at $z_{\rm
  abs}=0.3640$--0.6226, including one damped \lya\ absorber (DLA)
which also contains molecular hydrogen (see \S\ 4.1 below).  A suite
of ionic transitions is detected in these high $N(\HI)$ absorbers,
including low-ionization transitions such as \CII, \MgII, \SiII, and
\FeII, and intermediate-ionization transitions such as \CIII\ and/or
\NIII.  With the exception of the H$_2$-bearing DLA, the other four
LLSs also exhibit strong associated \OVI\ doublets (Figure 4).  The
observed relative column density ratios between low- and
intermediate-ionization transitions indicate that the gas is ionized.
Combining available galaxy survey data with known \HI\ absorber
properties provides new insights into the physical nature and origin
of optically-thick gas in galactic halos (e.g., Chen \etal\ 2005;
Kacprzak \etal\ 2010; Neeleman \etal\ 2016; P\'eroux \etal\ 2017;
Rudie \etal\ 2017; Chen \etal\ 2019a,b; P\'eroux \etal\ 2019;
Mackenzie \etal\ 2019; Lofthouse \etal\ 2020).  Here we summarize the
absorption properties of the optically-thick gas in descending order
of the observed $N(\HI)$, as well as the galactic environment
uncovered from available imaging and spectroscopic data.  A detailed
ionization analysis that accounts for the resolved component structure
in the ionic transitions is presented in Zahedy et al.\ (in
preparation).

\subsection{The DLA at $z_{\rm abs}=0.5762$ toward J0111$-$0316 near a massive, evolved galaxy at $d=42$ pkpc}

\subsubsection{Absorption properties of the optically-thick gas}

The absorber at $z_{\rm abs}=0.5762$ toward J0111$-$0316 is the
strongest \HI\ absorption system found in the LLS survey with
$\tau_{\rm 912}>6.6$.  This absorber turns out to be a DLA of
$\log\,N(\HI)/\cmjj=20.1_{-0.05}^{+0.15}$ with numerous features due
to the H$_2$ Lyman-Werner bands also detected in the COS spectrum.
The large $N(\HI)$ and the presence of H$_2$ indicate that the gas is
primarily neutral.  While low- and intermediate-ionization lines,
including \CII, \MgII, \SiII, and \CIII\ are present, \OVI\ is not
detected with a 2-$\sigma$ upper limit to the \OVI\ column density of
$\log\,N(\OVI)/\cmjj=13.6$.  The observed column densities of different
ions therefore provide a direct measurement of the underlying gas
metallicity, which we found to be $[{\rm M/H}]= -0.5\pm 0.2$, roughly
30\% of the solar value.  Details regarding this system, both the
absorption properties of the H$_2$-bearing DLA and its galactic
environment, are presented in Boettcher \etal\ (2020).

In summary, the \HI\ absorber exhibits a broad line width in the
high-order Lyman series lines that led to a best-fit $b_{\rm
  HI}\approx 49$ \kms.  While $N(\HI)$ is well constrained by damping
wings displayed in the \lyb\ profile, the broad $b_{\rm HI}$ is most
likely driven by the unresolved \HI\ component structure as indicated
by the associated metal lines which are resolved into three dominant
components at $d\,v_c=-60$, 0, $+70$ \kms\ (left column in Figure 4).
The components at $d\,v_c=-60$ and 0 \kms\ both exhibit associated
H$_2$ absorption with a best-fit H$_2$ column density of $\log\,N({\rm
  H_2})/\cmjj=15.6\pm 0.2$ and $19.0\pm 0.1$, respectively.  Because
the \HI\ component structure is not resolved at the velocity
separation between the two H$_2$ components, we calculate an
integrated mean H$_2$ fraction of $f_{\rm H_2}\equiv 2N({\rm
  H_2})/[N(\HI)+2\,N({\rm H_2})]=0.11\pm 0.02$.

\begin{table}
\scriptsize
\centering
\caption{Summary of best-fit Voigt profile parameters of new $\tau_{912}\apg 1$ absorbers$^a$}
\label{table:vp}
\begin{tabular}{lrrrr}
  \hline
  \hline
  \multicolumn{1}{c}{}	& \multicolumn{1}{c}{$dv_c$} & \multicolumn{1}{c}{} & \multicolumn{1}{c}{$b_c(\HI)$} & \multicolumn{1}{c}{} \\	
  \multicolumn{1}{c}{No.} & \multicolumn{1}{c}{(km/s)} & \multicolumn{1}{c}{log\,$N_c(\HI)/\mathrm{cm}^{-2}$} & \multicolumn{1}{c}{(km/s)} & \multicolumn{1}{c}{log\,$N_c(\MgII)/\mathrm{cm}^{-2}$}  \\ \hline 


  \multicolumn{5}{c}{J0111$-$0316\, $z_\mathrm{abs}=0.57616$, $\log\,N({\rm HI})/{\rm cm}^{-2}=20.1\pm0.1$$^b$}\\ \hline
1	&	$-186$	& $15.8\pm 0.05$        & $15^{+1}_{-2}$ & $<11.1$ \\  
2	&	$-16$	& $20.10^{+0.15}_{-0.05}$ & $49^{+0}_{-1}$ & $>14.0$ \\  
3	&	$+178$	& $14.9\pm0.1$		& $11^{+4}_{-3}$ & $<11.1$ \\  \hline
\multicolumn{5}{c}{J2135$-$5316, $z_\mathrm{abs}=0.62255$, $\log\,N({\rm HI})/{\rm cm}^{-2}=18.01\pm0.04$}\\ \hline
1	&	$-268.3^{+2.3}_{-2.6}$	& $14.51^{+0.03}_{-0.04}$& $32.5^{+2.7}_{-2.6}$  & $<11.0$ \\  
2	&	$-119.7^{+0.5}_{-0.7}$	& $15.38\!\pm\!0.02$		& $26.8\!\pm\!0.6$ & $<11.0$ \\  
3	&	$0.0$				& $18.00\!\pm\!0.04$		& $18.7\!\pm\!0.2$ & $12.76\pm 0.01$  \\  
4	&	$+26.4$				& $15.87^{+0.10}_{-0.38}$& $18.5^{+3.7}_{-2.8}$ & $12.53\pm 0.01$ \\  
5	&	$+55.5^{+3.6}_{-8.4}$	& $15.32^{+0.22}_{-0.11}$	& $21.9^{+4.5}_{-1.8}$  & $<11.0$ \\  
6	&	$+98.0^{+14.3}_{-13.4}$	& $14.17^{+0.10}_{-0.16}$& $83.0^{+17.9}_{-14.3}$ & $<11.0$ \\    \hline
\multicolumn{5}{c}{J2308$-$5258, $z_\mathrm{abs}=0.54273$, $\log\,N({\rm HI})/{\rm cm}^{-2}=17.59\pm0.02$}\\ \hline
1	&	$-144.4$	& $16.12\!\pm\!0.03$		& $12.2^{+0.8}_{-0.6}$ & $12.96\pm 0.02$ \\  
2	&	$-114.3\pm3.1$	& $15.44\!\pm\!0.02$		& $59.6^{+3.3}_{-3.0}$ & $<11.1$ \\  
3	&	$-23.9$		& $16.67^{+0.03}_{-0.05}$ & $16.1^{+0.6}_{-0.5}$ & $ 12.98\pm 0.04$ \\  
4	&	$0.0$		& $17.46\!\pm\!0.02$		& $14.4^{+1.0}_{-0.7}$ & $13.46\pm 0.05$ \\  
5	&	$+23.9$		& $16.53^{+0.06}_{-0.10}$ & $18^c$ & $13.13\pm 0.10$ \\
6	&	$+42.3$		& $15.90\!\pm\!0.08$		& $15.7^{+1.5}_{-1.2}$ & $12.90\pm 0.01$ \\  \hline
\multicolumn{5}{c}{J0248$-$4048, $z_\mathrm{abs}=0.36400$, $\log\,N({\rm HI})/{\rm cm}^{-2}=17.57\pm0.01$}\\ \hline
1	&	$-130.8^{+1.2}_{-1.0}$	& $15.42\!\pm\!0.02$		& $31.1^{+1.2}_{-0.8}$ & $<11.3$ \\  
2	&	$-42.1$		        & $16.64\!\pm\!0.02$		& $23.8^{+0.5}_{-0.7}$ & $12.55\pm 0.03$ \\  
3	&	$0.0$			& $17.51\!\pm\!0.01$		& $20.0\!\pm\!0.3$ & $13.44\pm 0.03$ \\  
4	&	$+57.3^{+1.2}_{-1.6}$ 	& $15.36\!\pm\!0.04$		& $9.3^{+1.5}_{-1.0}$  & $<11.3$ \\  
5	&	$+100.0^{+8.4}_{-10.1}$ 	& $14.70\!\pm\!0.08$		& $69.0^{+6.1}_{-5.6}$ & $<11.3$ \\  
6	&	$+154.4^{+1.5}_{-1.6}$ 	& $14.59\!\pm\!0.05$		& $19.2^{+2.3}_{-1.7}$ & $<11.3$ \\  \hline
\multicolumn{5}{c}{J0357$-$4812, $z_\mathrm{abs}=0.43527$, $\log\,N({\rm HI})/{\rm cm}^{-2}=17.18\pm0.01$}\\ \hline
1	&	$-35.5$				& $16.37\!\pm\!0.02$		& $24.2^{+0.4}_{-0.3}$ & $11.47\pm 0.16$ \\  
2	&	$0.0$				& $17.07\!\pm\!0.01$		& $17.4\!\pm\!0.5$ & $12.10\pm 0.05$ \\  
3	&	$+31.4$				& $16.07\!\pm\!0.03$		& $18.8\!\pm\!0.5$ & $12.38\pm 0.11$ \\  
4	&	$+85.8\pm3.5$			& $13.55^{+0.10}_{-0.14}$& $13.4\!\pm\!2.8$ & $<11.3$ \\  
5	&	$+133.1^{+3.7}_{-4.9}$	& $13.60^{+0.07}_{-0.06}$& $27.9^{+3.7}_{-2.8}$ & $<11.3$ \\  
6	&	$+251.7^{+2.1}_{-1.7}$	& $13.89^{+0.07}_{-0.06}$& $16.2^{+2.1}_{-2.0}$ & $<11.3$ \\  \hline
\multicolumn{5}{l}{$^a$Best-fit velocity centroid $dv_c$ of individual components relative to $z_{\rm abs}$, \HI}\\
\multicolumn{5}{l}{\ \ component column density $N_c(\HI)$ and Doppler parameter $b_c$, and \MgII}\\
\multicolumn{5}{l}{\ \ component column density $N_c(\MgII)$ from Zahedy \etal\ (2020).}\\
\multicolumn{5}{l}{$^b$Because the primary \HI\ components are not resolved, $N_c(\MgII)$ here} \\
\multicolumn{5}{l}{\ \ represents the sum of all seven resolved components in the optical echelle}\\
\multicolumn{5}{l}{\ \ spectrum and $d\,v_c=0$ \kms\ corresponds to the strongest \MgII\ component.}\\
\multicolumn{5}{l}{\ \ See Boettcher et al.\ (2020) for details.  }\\
\multicolumn{5}{l}{$^c$Due to heavy blending with adjacent \HI\ components, the $b$ value is fixed}\\
\multicolumn{5}{l}{\ \ to the expected thermal value for the temperature determined from the}\\
\multicolumn{5}{l}{\ \ corresponding \MgII\ component.} \\
\end{tabular}
\end{table}

\subsubsection{The galaxy environment}

A group of nine galaxies of $r$-band magnitude $AB(r)=20.4-24$ mag
has been spectroscopically identified with Magellan at projected
distance $d<600$ pkpc and line-of-sight velocity interval $d\,v_{\rm
  gal}< 300$ \kms\ from the H$_2$-bearing DLA.  The closest galaxy is
identified at $d=42$ pkpc and $d\,v_{\rm gal}=-57$ \kms\ from the DLA
and it is bright with $AB(r)=21.47$, corresponding to $\approx
1.9\,L_*$ at $z=0.576$ (see e.g., Cool \etal\ 2012).  Both the
broad-band photometric colors and the optical spectrum of the closest
galaxy indicate that it is an evolved galaxy of stellar mass $M_{\rm
  star}=8\times 10^{10}\,\msun$ and no trace of on-going star
formation with a 2-$\sigma$ upper limit on the star formation rate
(SFR) of ${\rm SFR}<0.2\,\msun\,{\rm yr}^{-1}$.  Two more galaxies are
found at $d<300$ pkpc.  Both are sub-$L_*$ galaxies with
$\log\,\mstar/\msun=9.4$ and 9.9 at $d=130$ and 167 pkpc,
respectively.
This represents one of two H$_2$ absorbers found in the vicinity of a
massive, evolved galaxy beyond the nearby universe (see Zahedy
\etal\ 2020 for a second case, and Muzahid \etal\ 2015, 2016 for a
list of H$_2$ absorbers found in the vicinities of star-forming
galaxies).

\begin{figure*}
\includegraphics[scale=0.45]{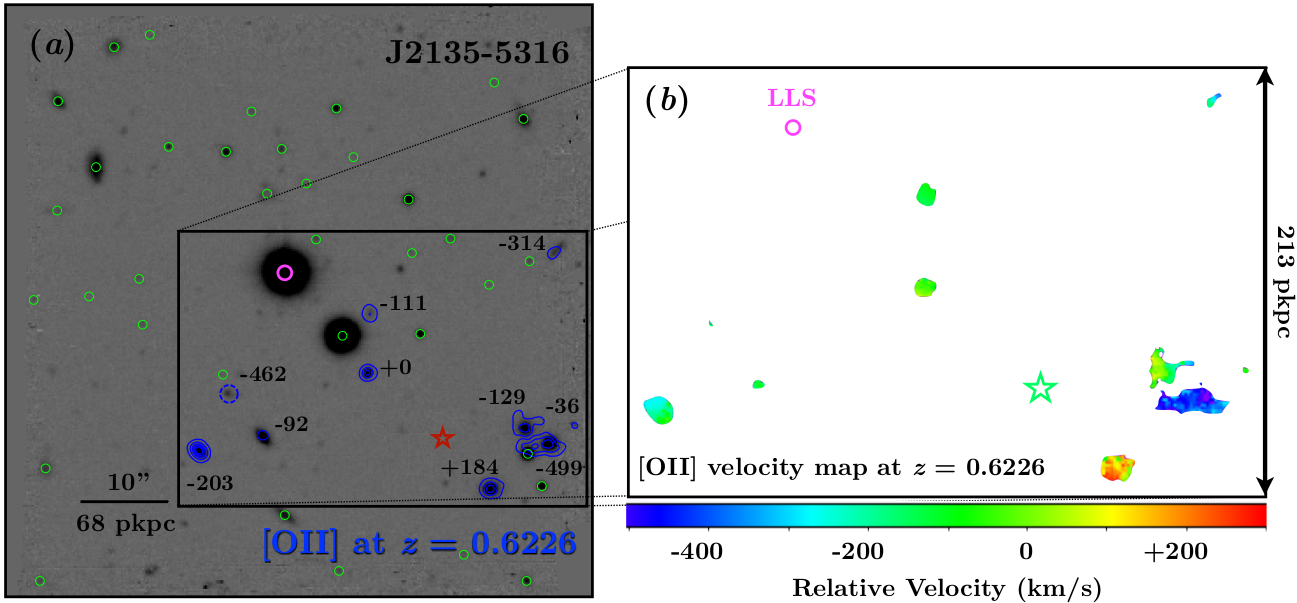}
\caption{The galaxy environment uncovered by MUSE for the LLS at
  $z_{\rm abs}=0.6226$ toward J2135$-$5316.  The pseudo $r$-band image
  from Figure 2 is reproduced in panel ({\it a}).  North is up and
  east is to the left.  Objects that are spectroscopically identified
  at cosmologically distinct redshifts from the LLS are marked by
  green circles, while galaxies spectroscopically identified in the
  vicinity of the LLS are highlighted by their blue [\OII] emission
  contours of constant surface brightness 0.25, 1.25, 2.5, 3.7, and
  $5\times 10^{-17}\,{\rm erg}\,{\rm s}^{-1}\,{\rm cm}^{-2}\,{\rm
    arcsec}^{-2}$ and marked by their line-of-sight velocity offset (\kms)
  from $z_{\rm abs}=0.6226$.  The galaxy at $\Delta\,v=-462$
  \kms\ exhibits only a faint trace of [\OII] emission, with surface
  flux density below the lowest contour.  It is marked by a blue,
  dashed circle.  A group of 10 galaxies is found in the vicinity of
  the LLS with angular distance ranging from $\theta=10.6''$ to
  $38.1''$ from the QSO sightline (magenta circle), corresponding to a
  range in projected distance from $d=72$ to 252 pkpc at the redshift
  of the LLS.  The light-weighted center of the galaxy group is marked
  by an open star symbol at ($-17.9''$, $-19.0''$) from the QSO
  sightline.  The line-of-sight velocity map of [\OII] emission is
  presented in the right panel with zero velocity corresponding to
  $z_{\rm abs}=0.6226$.  The location of the LLS is marked by an open
  magenta circle.  The galaxy group spans a range in the line-of-sight
  velocity offset from $d\,v_{\rm gal}=-499$ \kms\ to $d\,v_{\rm
    gal}=+184$ \kms\ from the absorber with a light-weighted center at
  $d\,v_{\rm gal}=-170$ \kms\ (open star symbol in panel {\it b}).
  While the galaxy group members all exhibit an ordinary continuum
  morphology in the pseudo $r$-band image, the [\OII] contours
  revealed spatially-extended line-emitting nebulae around the two
  massive group members at $d\,v_{\rm gal}=-129$ and $-499$ \kms\ in
  the lower-right corner, suggesting strong interactions between the
  two galaxies.  The galaxy at $d\,v_{\rm gal}=-499$ \kms\ is also the
  most massive member of the group with spectral features indicative
  of a post-starburst phase (and possibly hosting an AGN; see text for
  details).}
\label{figure:J2135gmap}
\end{figure*}

\subsection{The LLS at $z_{\rm abs}=0.6226$ toward J2135$-$5316 in a massive galaxy group at $d_{\rm group}=177$ pkpc}

\subsubsection{Absorption properties of the optically-thick gas}

The LLS at $z_{\rm abs}=0.6226$ toward J2135$-$5316 has $\tau_{\rm
  912}>6.3$.  It is the second strongest LLS found in our sample
(after the H$_2$-bearing DLA toward J0111$-$0316), and is resolved
into a minimum of six components with $\approx 99$\% of the $N(\HI)$
contained in the central component at $d\,v_c=0$ \kms\ (component 3 in
Table 6 for this system) and $\approx 1$\% contained in the components
at $d\,v_c=-120$, $+26$, and $+56$ \kms\ (components 2, 4, and 5,
respectively, in Table 6).  The remaining components contribute
negligibly to the total $N(\HI)$, but they dominate the line width
observed in the first few Lyman lines at $d\,v_c=-268$ \kms\ and
$d\,v_c=+98$ \kms\ (Figures 3 and 4).  In particular, the large $b$
value of component 6 ($b_c\approx 83$ \kms\ and
$\log\,N_c(\HI)/\cmjj\approx 14.2$) in Table 6 is likely due to blended
weak \HI\ components that are not resolved in the COS spectra, but are
necessary for explaining the observed \lya\ and \lyb\ line profiles.
This component is not present in higher-order Lyman lines and the lack
of resolving power has a negligible impact on the total $N(\HI)$
measurement, but the best-fit $N_c(\HI)$ for component 6 should be
taken with caution.

While low-ionization species are
concentrated in the two strongest \HI\ components at $d\,v_c=0$ and
$+26$ \kms\ (components 3 \& 4 in Table 6), intermediate ions exhibit
matching component structure with the first few Lyman series lines at
$-120\apl d\,v_c\apl 60$ \kms.
In particular, component 2 at $d\,v_c=-120$
\kms\ exhibits strong and relatively narrrow \CIII\ and \OVI\ absorption and no detectable
low-ionization transitions, showing a distinct ionization
state from components 3 \& 4).

\subsubsection{The galaxy environment}

The MUSE observations of this field presented in \S\ 2.4 have
uncovered 86 objects in the $1'\times 1'$ field of view with pseudo
$r$-band magnitude ranging from $AB(r)=21$ to 26.6 mag, in addition to
a bright star and the QSO.  Robust redshift measurements are obtained
for 46 of these objects ({\it left} panel of Figure 5).  Comparing the
photometric measurements of common objects observed in both MUSE and
DES shows that the pseudo $r$-band magnitudes are consistent with DES
$r$-band measurements to within 0.1 magnitude uncertainties.  The
spectroscopic survey is 100\%, $>90$\%, and 75\% complete for objects
brighter than $AB(r)=23$, 24, and 25 mag, respectively.  Redshift
measurements are made using a $\chi^2$ fitting routine that compares
the observed spectrum with models formed from a linear combination of
four eigenspectra from the SDSS at different redshifts (see Chen \&
Mulchaey 2009 and Johnson \etal\ 2013 for a detailed description).
The best-fit redshift of each object was visually inspected for
confirmation.  The redshifts of spectroscopically identified galaxies
range from $z_{\rm gal}=0.24$ to $z=4.49$.

\begin{figure}
\includegraphics[scale=0.45]{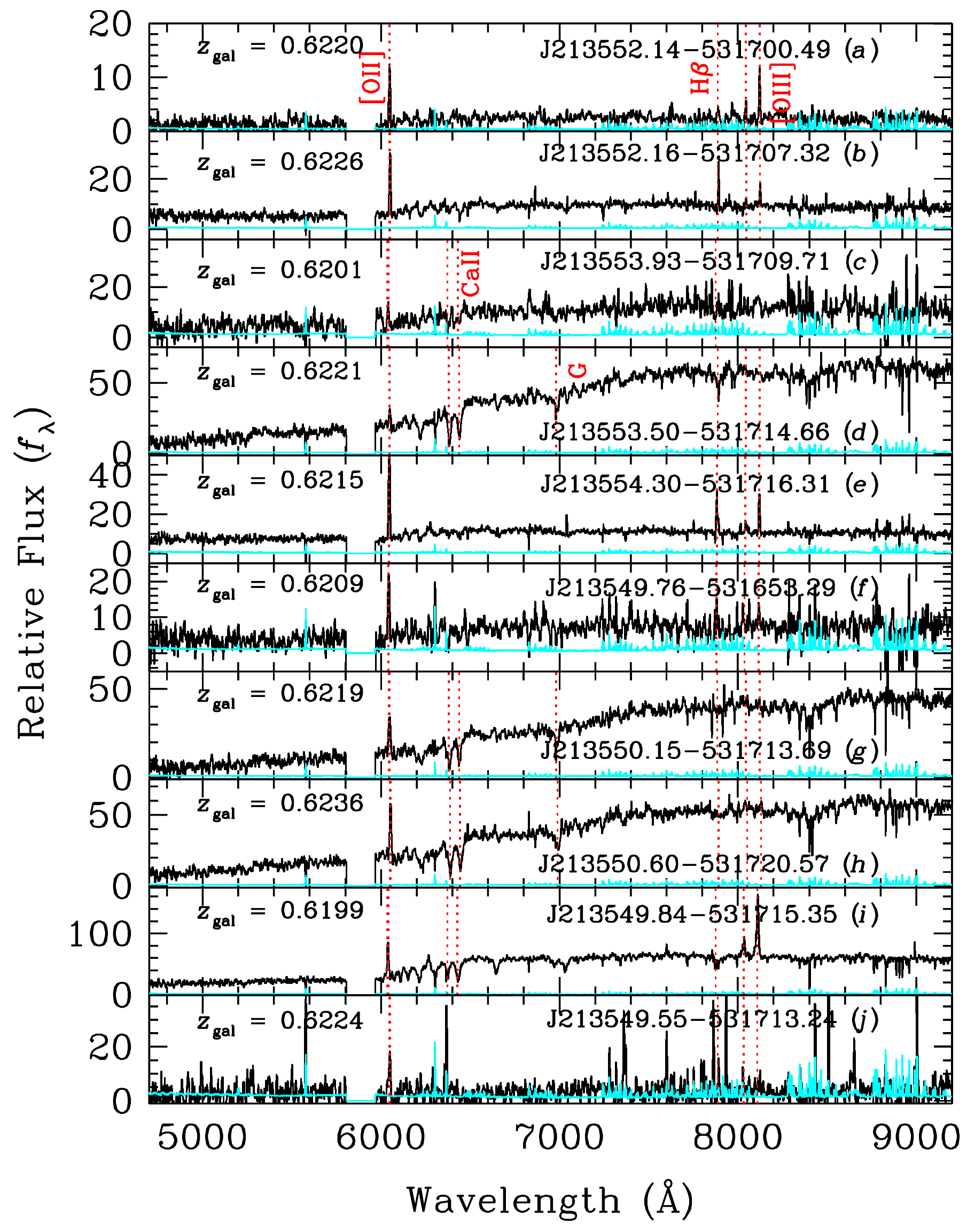}
\caption{Optical spectra of a group of 10 galaxies identified in the
  vicinity of the LLS at $z_{\rm abs}=0.6226$ toward J2135$-$5258 from
  MUSE observations.  The corresponding 1-$\sigma$ error spectra are
  shown in cyan.  The spectra are ordered with increasing projected
  distance of the galaxies from $d=72$ pkpc at the top to $d=252$ pkpc
  at the bottom.  Galaxies in the group exhibit a wide range in star
  formation history with the most luminous and massive members
  displaying prominent absorption features such as \CaII\ and G-band
  (panels {\it d}, {\it g}, and {\it h}), indicative of evolved old
  stars, one galaxy displaying a strong Balmer absorption series with
  weak H$\beta$ and [\OII] (panel {\it i}), indicative of a post
  starburst phase, and the rest displaying strong [\OII] emission and
  no significant \CaII\ absorption (panels {\it a}, {\it b}, {\it e},
  {\it f}, and {\it j}), indicating a predominantly young stellar
  population.  The galaxy at $z_{\rm gal}=0.6199$ (panel {\it i}) also
  exhibits [\NeIII] and a high [\OIII]/H$\beta$ line ratio, suggesting
  the presence of an AGN.}
\label{figure:J2135gspec}
\end{figure}

A group of 10 galaxies is found in the vicinity of the LLS with pseudo
$r$-band magnitude ranging from $AB(r)=21.2$ to 24.7 mag and
line-of-sight velocity offset ranging from $d\,v_{\rm gal}=-499$
\kms\ to $d\,v_{\rm gal}=+184$ \kms\ (Figure 5).  The remaining
spectroscopically-identified galaxies in the MUSE field all appear at
a cosmologically distinct redshift with a velocity separation
exceeding $|d\,v_{\rm gal}|=600$ \kms\ from the LLS.  The galaxies in
the LLS-associated group span a range in angular distance from
$\theta=10.6''$ to $\theta=38.1''$, corresponding to a range in
projected distance from $d=72$ to 252 pkpc at the redshift of the LLS.
Optical spectra of the members of the LLS-associated galaxy group are
presented in Figure 6 with increasing projected distance from top to
bottom, which clearly show a wide range in the star formation
histories among the members of the galaxy group.  A detailed analysis
of galaxy properties is presented in Appendix A.  In summary, we
estimate its intrinsic $r$-band absolute magnitude $M_r$ based on the
observed $AB(r)$ and find that the group galaxies span a range in
$M_r$ from $M_r=-22.5$ (or $3\,L_*$) for a post starburst/AGN at
$d=244$ pkpc (see Appendix A) to $M_r=-18.3$ (or $0.06\,L_*$) for a
faint dwarf at $d=252$ pkpc and $d\,v_{\rm gal}=-36$ \kms.  In
addition, we estimate \mstar\ based on the rest-frame $g-r$ color
inferred from the MUSE spectra and the prescription of Johnson
\etal\ (2015), and find $\log\,\mstar/\msun\,=8.6-11.2$ for these
galaxies.  The observed and derived properties of the galaxies,
including best-fit redshift $z_{\rm gal}$, projected distance $d$,
mean line-of-sight velocity offset $|d\,v_{\rm gal}|$, pseudo $r$-band
magnitude $AB(r)$, intrinsic absolute $r$-band magnitude $M_r$,
inferred total stellar mass \mstar, galaxy type, The total integrated
[\OII] line flux, and unobscured SFR are summarized in columns (2)
through (10) of Table 7.

The available MUSE data show that while the closest galaxy to the LLS
is a $0.12\,L_*$ galaxy with $\mstar\approx 7.9\times 10^8\,\msun$ at
$d=72$ pkpc and $d\,v_{\rm gal}=-111$ \kms, it is likely part of a
dynamic group containing several members that are 100 times more
massive at $<200$ pkpc away.  All 10 galaxies contribute to a total
stellar mass of $\mstar\approx 3.4\times 10^{11}\,\msun$ with $\approx
60$\% contained in the interacting pair.  We calculate a
light-weighted center located at ($-17.9''$, $-19.0''$) from the QSO
sightline (open star symbol in Figure 5).  The corresponding projected
distance of the `group' is therefore $d_{\rm group}=177$ pkpc and the
light-weighted line-of-sight velocity offset is $d\,v_{\rm
  group}\approx -170$ \kms\ between the galaxy `group' and the LLS.
We also calculate a line-of-sight velocity dispersion of $\sigma_{\rm
  group}=211$ \kms, implying a dynamical mass of $M_{\rm dyn}\approx
1.1\times 10^{13}\,\msun$ and a virial radius of $r_{\rm vir}\approx
360$ kpc for the galaxy group.  The strong LLS may be due to gaseous
streams at $\apg 0.5\,r_{\rm vir}$ in the intragroup medium (e.g.,
Chen \etal\ 2019b).

\subsection{The LLS at $z_{\rm abs}=0.5427$ toward J2308$-$5258 near a star-forming galaxy at $d=32$ pkpc}

\subsubsection{Absorption properties of the optically-thick gas}

The LLS at $z_{\rm abs}=0.5427$ toward J2308$-$5258 has $\tau_{\rm
  912}=2.53\pm 0.01$.  It is resolved into a minimum of six components
with $\approx 74$\% of $N(\HI)$ contained in the central component
(component 4 in Table 6) at $d\,v_c=0$ \kms\ and $\approx 21$\%
contained in the two components at $d\,v_c=-24$ and $+24$
\kms\ (components 3 and 5, respectively, in Table 6).  Different from
previous systems, the remaining components, in particular components 1
and 6, contribute an appreciable amount (3\% and 2\%, respectively) to
the total $N(\HI)$ (Figures 3 and 4).  However, the large $b$ value of
component 2 ($b_c\approx 60$ \kms\ and $\log\,N_c(\HI)/\cmjj\approx 15.4$)
in Table 6 is likely due to unresolved \HI\ components in the COS
spectra, but is necessary for explaining the observed \lya\ and
\lyb\ line profiles.  Because this component is not present in
higher-order Lyman lines, the lack of resolving power has a negligible
impact ($<1$\%) on the total $N(\HI)$ measurement but the best-fit
$N_c(\HI)$ for component 2 should be taken with caution.

\begin{table*}
\scriptsize
\centering
\caption{Intrinsic properties of galaxies in the vicinity of a LLS}
\label{table:group}
\centering {
\begin{tabular}{lrrrccrcccc}
\hline \hline
                           &                          & \multicolumn{1}{c}{$d$}   & \multicolumn{1}{c}{$d\,v_{\rm gal}^a$} &                           &                          &                         &                         & \multicolumn{1}{c}{$f_{\rm line}$$^d$} &   \multicolumn{1}{c}{SFR$^e$}    &   \multicolumn{1}{c}{ISM} \\
\multicolumn{1}{c}{Galaxy} & \multicolumn{1}{c}{$z_{\rm gal}$}    & \multicolumn{1}{c}{(pkpc)} & \multicolumn{1}{c}{(km/s)}       & \multicolumn{1}{c}{$AB(r)$$$} & \multicolumn{1}{c}{$M_r$$^b$} & \multicolumn{1}{c}{$\log\,M_{\rm star}/{\rm M}_\odot$} & \multicolumn{1}{c}{Type$^c$} & \multicolumn{1}{c}{($10^{-17}$\,erg/s/cm$^{2}$)} & \multicolumn{1}{c}{(${\rm M}_\odot/{\rm yr}$)} & \multicolumn{1}{c}{(O/H)$^f$} \\
\multicolumn{1}{c}{(1)}    & \multicolumn{1}{c}{(2)}  & \multicolumn{1}{c}{(3)}   & \multicolumn{1}{c}{(4)}         & \multicolumn{1}{c}{(5)}   & \multicolumn{1}{c}{(6)}  & \multicolumn{1}{c}{(7)} & \multicolumn{1}{c}{(8)} & \multicolumn{1}{c}{(9)} & \multicolumn{1}{c}{(10)} & \multicolumn{1}{c}{(11)} \\
\hline
\multicolumn{10}{c}{Galaxies associated with the LLS at $z_{\rm abs}=0.6226$ toward J2135$-$5258} \\
\hline
J213552.14$-$531700.49 & 0.6220 &  72 & $-$111 & 24.0 & $-19.0$  &  8.9  &  SF  &  $2.06\pm 0.03$  & 0.22  & ...  \\
J213552.16$-$531707.32 & 0.6226 & 100 &     0  & 22.6 & $-20.4$  &  9.4  &  SF  &  $3.77\pm 0.04$  & 0.40  & ...  \\
J213553.93$-$531709.71 & 0.6201 & 104 & $-$462 & 23.7 & $-20.0$  & 10.2  &   A  &  $0.05\pm 0.01$  & 0.01  & ...  \\
J213553.50$-$531714.66 & 0.6221 & 129 &  $-$92 & 21.6 & $-22.0$  & 10.7  &   A  &  $0.64\pm 0.08$  & 0.07  & ...  \\
J213554.30$-$531716.31 & 0.6215 & 154 & $-$203 & 22.3 & $-20.7$  &  9.6  &  SF  &  $1.14\pm 0.05$  & 0.12  & ...  \\
J213549.76$-$531653.29 & 0.6209 & 211 & $-$314 & 23.0 & $-20.0$  &  9.3  &  SF  &  $1.00\pm 0.03$  & 0.11  & ...  \\
J213550.15$-$531713.69 & 0.6219 & 222 & $-$129 & 21.8 & $-21.9$  & 10.6  &   A  &  $4.96\pm 0.16$  & 0.53  & ...  \\
J213550.60$-$531720.57 & 0.6236 & 231 & $+$184 & 21.5 & $-22.2$  & 10.8  &   A  &  $6.47\pm 0.30$  & 0.69  & ...  \\
J213549.84$-$531715.35 & 0.6199 & 244 & $-$499 & 21.2 & $-22.5$  & 11.2  & PSB  & $15.23\pm 0.72$  & 1.63  & ...  \\
J213549.55$-$531713.24 & 0.6224 & 252 &  $-$36 & 24.7 & $-18.3$  &  8.6  &  SF  &  $0.26\pm 0.02$  & 0.03  & ...  \\
\hline
\multicolumn{10}{c}{Galaxy associated with the LLS at $z_{\rm abs}=0.5427$ toward J2308$-$5258$^g$} \\
\hline
J230837.42$-$525845.07  & 0.5426 &  32 & $-19$ & 21.0 & $-22.1$  & 10.9 &  SF$+$A    & ... & ... & ...  \\
\hline
\multicolumn{10}{c}{Galaxies associated with the LLS at $z_{\rm abs}=0.3640$ toward J0248$-$4048} \\
\hline
J024806.09$-$404835.64  & 0.3638 &  15 & $-$44 &  ...  &   ...    & ... &   SF    & $4.41\pm 0.03$ & 0.11 & $7.7$          \\
J024805.69$-$404831.04  & 0.3639 &  37 & $-$22 &  25.1 & $-16.3$  & 7.8 &   SF    & $0.63\pm 0.02$ & 0.01 & $7.5$  \\
J024803.87$-$404822.24  & 0.3637 & 150 & $-$66 &  24.1 & $-17.3$  & 8.2 &   SF    & $1.98\pm 0.02$ & 0.05 & $7.6$  \\
\hline
\multicolumn{10}{c}{Galaxy associated with the LLS at $z_{\rm abs}=0.4353$ toward J0357$-$4812} \\
\hline
J035723.07$-$481215.78  & 0.4358 &  67 & $+105$ & 20.7 & $-21.7$  & 10.4 &  SF$+$A    & $6.14\pm 0.06$ & 0.28 & ...  \\
\hline
\multicolumn{11}{l}{$^\mathrm{a}$Line-of-sight velocity offset relative to the strongest \HI\ absorbing component at $z_{\rm abs}$.} \\
\multicolumn{11}{l}{$^\mathrm{b}$At $z=0.3$, star-forming galaxies (SF) have a characteristic rest-frame absolute $r$-band magnitude of $M_{r_*}=-21.3$ and evolved galaxies (A) have} \\
\multicolumn{11}{l}{\ \ $M_{r_*}=-21.5$ (e.g., Cool \etal\ 2012).} \\
\multicolumn{11}{l}{$^\mathrm{c}$SF: emission-line dominated star-forming; PSB: post starburst; A: absorption-dominated evolved galaxies.} \\
\multicolumn{11}{l}{$^\mathrm{d}$Integrated line flux for H$\alpha$ at $z<0.4$ and [\OII] at higher redshifts.} \\
\multicolumn{11}{l}{$^\mathrm{e}$Unobscured SFR inferred from the integrated line flux reported in column (9) without dust extinction corrections.  The conversion from H$\alpha$ is based}\\
\multicolumn{11}{l}{\ \ on the calibrator of Kennicutt \&  Evans (2012), while the conversion from [\OII] is taken from Kewley \etal\ (2004).  For galaxies at $z<0.4$, the} \\
\multicolumn{11}{l}{\ \ observed H$\alpha$/H$\beta$ ratio indicates that the dust-extinction corrected SFR should be about between 24\% and 100\% higher (see \S\ 4.2 for discussion).}\\
\multicolumn{11}{l}{$^\mathrm{f}$Uncertainties are driven by the $R_{23}$ calibration error and are estimated to be about 0.1 dex.  For comparison, the oxygen abundance of the Sun is} \\
\multicolumn{11}{l}{\ \ $12 + \log({\rm O}/{\rm H})_\odot = 8.69\pm 0.05$ (Asplund \etal\ 2009).} \\
\multicolumn{11}{l}{$^\mathrm{g}$MUSE observations are on-going.  Available partial data are sufficient for identifying a luminous galaxy in the vicinity of the LLS.  A full analysis is}\\
\multicolumn{11}{l}{\ \ deferred to a later paper when the observations have been completed.} \\
\end{tabular}
}
\end{table*}

There is a good one-to-one correspondence between
the associated low-ionization transitions and the \HI\ components with
the only exception being component 2, the broad component at
$d\,v_c\approx -114$ \kms.  The kinematic profiles of low ions also
match well with those of intermediate- and highly-ionized species.
However, it is clear that these components exhibit distinct
$N(\HI)$ to $N(\MgII)$ ratios.  
Specifically, components 1
and 3 exhibit a column density ratio of 2:7 in $N(\HI)$, while the
associated \MgII\ components exhibit a comparable strength.  The
differential $N(\MgII)/N(\HI)$ ratios strongly imply differences in
the underlying gas metallicity between these components (e.g., Zahedy
\etal\ 2019).
Different from the LLS at $z=0.6226$ toward J2135$-$5316, the \OVI\ doublet of this LLS
is broad and covers the velocity range of all lower-ionization
species, highlighting the multiphase nature of the gas.

\begin{figure}
\includegraphics[scale=0.35]{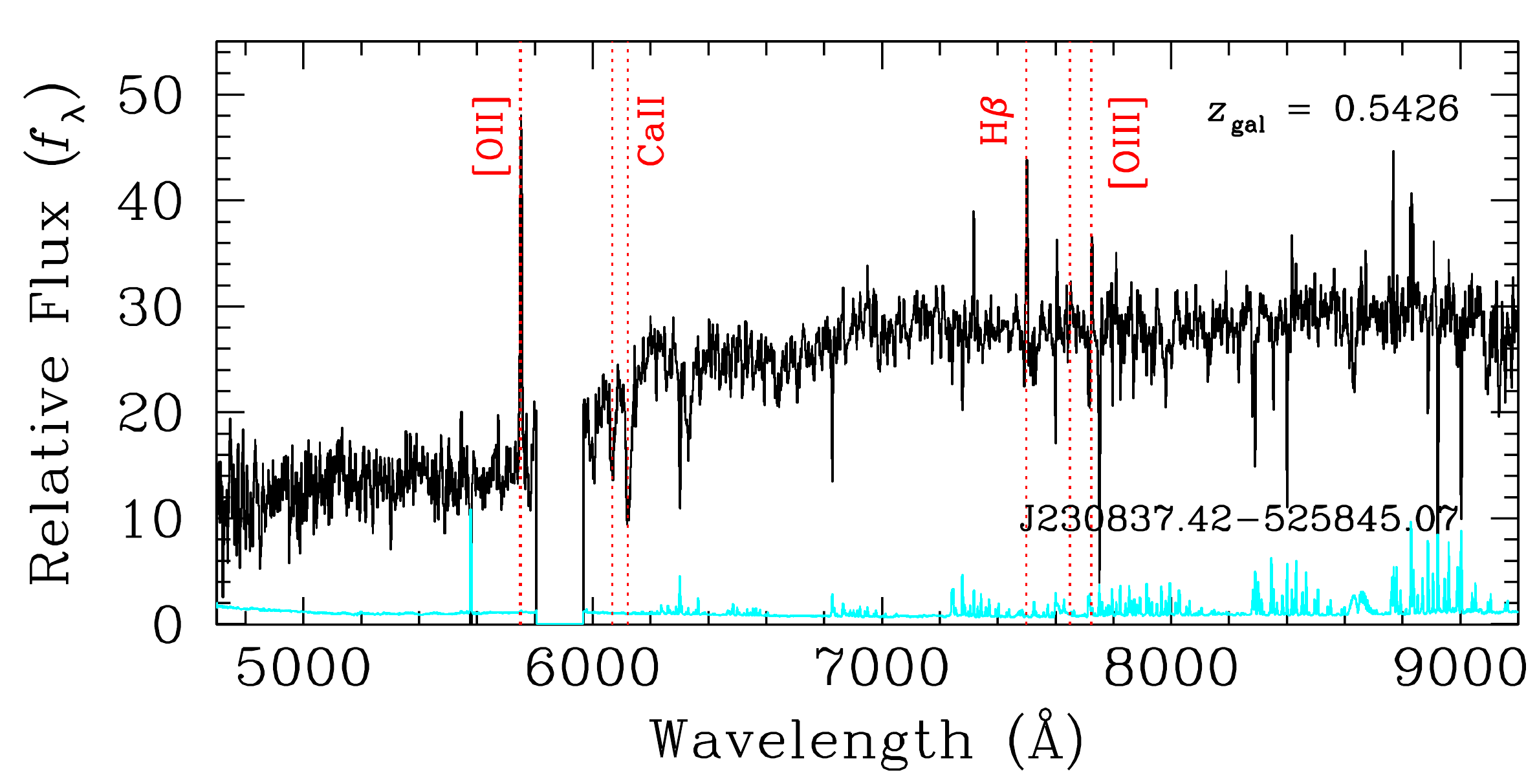}
\caption{Optical spectrum of a luminous galaxy identified in the
  vicinity of the LLS at $z_{\rm abs}=0.5427$ toward J2308$-$5258 from
  available MUSE data.  The corresponding 1-$\sigma$ error spectrum is
  shown in cyan.  The galaxy is bright and the spectrum exhibits
  prominent absorption features due to \CaII\ H\&K and the Balmer
  series, along with strong [\OII] and H$\beta$ emission lines
  (highlighted in red dotted lines with corresponding line
  identifications). }
\label{figure:J2135gspec}
\end{figure}

\subsubsection{The galaxy environment}

\begin{figure*}
\includegraphics[scale=0.5]{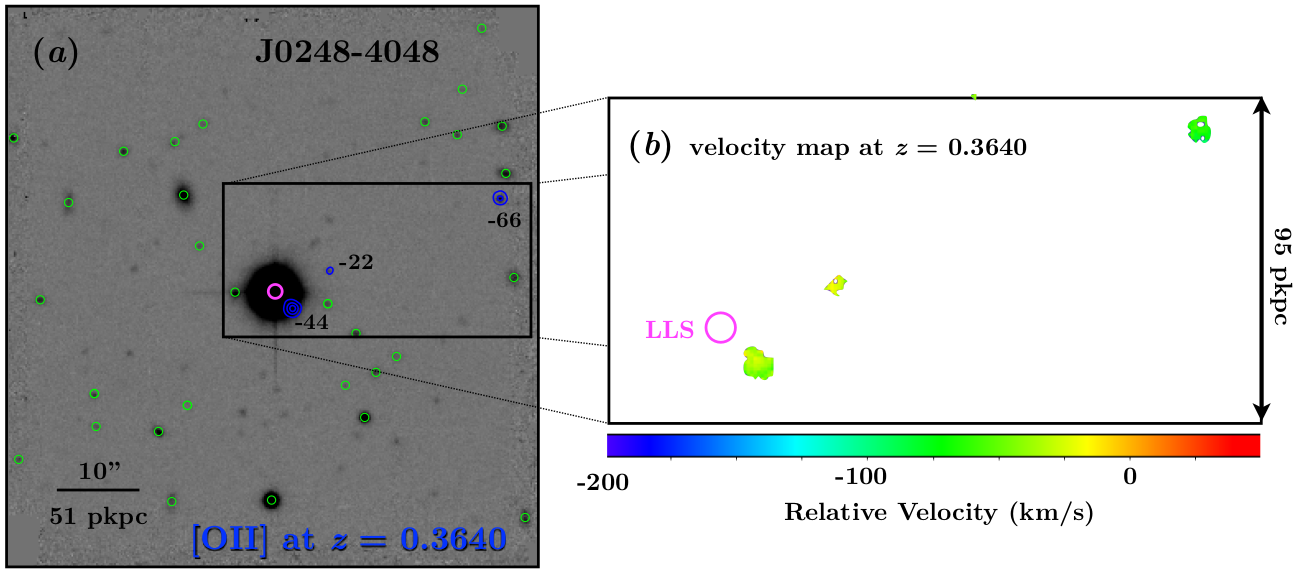}
\caption{The galaxy environment uncovered by MUSE for the LLS at
  $z_{\rm abs}=0.3640$ toward J0248$-$4048.  The pseudo $r$-band image
  from Figure 2 is reproduced in panel ({\it a}).  North is up and
  east is to the left.  Objects that are spectroscopically identified
  at a cosmologically distinct redshift from the LLS are marked by
  green circles, while galaxies associated with the LLS are
  highlighted by blue [\OII] emission contours of constant surface
  brightness 0.25, 1.25, and $2.5\times 10^{-17}\,{\rm erg}\,{\rm
    s}^{-1}\,{\rm cm}^{-2}\,{\rm arcsec}^{-2}$ and marked by the
  line-of-sight velocity offset (\kms) from $z_{\rm abs}=0.3640$.  Three
  galaxies are found in the vicinity of the LLS with angular distance
  of $\theta=3.0''$, $7.8''$, and $29.7''$ from the QSO sightline
  (magenta circle), corresponding to $d=15$, 37, and 150 pkpc,
  respectively, at the redshift of the LLS.  The line-of-sight
  velocity map based on a simultaneous fit to the observed [\OII]
  doublet, H$\beta$, [$\OIII$]$\lambda\lambda\,4960,5008$, and
  H$\alpha$ (see Figure 9) is presented in the right panel with zero
  velocity corresponding to $z_{\rm abs}=0.3640$.  The location of the
  LLS is marked by an open magenta circle.  All three LLS-associated
  galaxies are found at relatively small line-of-sight velocity offset
  of $|\Delta\,v|<70$ \kms\ from the absorber.  }
\label{figure:J0248gmap}
\end{figure*}

The MUSE observations of this field are ongoing.  Only a third of the
observations have been executed.  It is expected that the data will
reach a comparable depth as seen in J0357$-$4812 (see \S\ 4.5 below).
An initial analysis of the data collected so far has revealed a
luminous disk galaxy of $AB(r)\approx 21$ mag at $z_{\rm gal}=0.5426$
and $\theta=5.1''$ from the QSO sightline, corresponding to $d=32$
pkpc and $d\,v_{\rm gal}=-19$ \kms\ from the LLS.  At $z=0.54$, the
observed pseudo $r$-band magnitude implies $M_r\approx -22.1$ (or
$2.1\,L_*$) and $\log\,\mstar/\msun\approx 10.9$.  Similar to the
LLS-absorbing galaxy in J0357$-$4812 (see \S\ 4.5 and Figure 11), the
optical spectrum of this galaxy is characterized by a combination of
strong \CaII\, H\&K absorption and strong [\OII] and H$\beta$ emission
lines (Figure 7).  Different from the configuration between the QSO
probe and the LLS absorbing galaxy in J0357$-$4812, the QSO here
probes the diffuse gas at $\approx 28^\circ$ from the minor axis of
the inclined disk.  While no other galaxies are found in the vicinity
of the LLS to $AB(r)\approx 25.5$ mag, suggesting that the luminous
galaxy at $d=32$ pkpc is singularly responsible for the LLS, the
survey is still incomplete.  A detailed analysis of the
galactic environment of this LLS will be presented in a later paper
when complete galaxy survey data are available from MUSE and Magellan.

\subsection{The LLS at $z_{\rm abs}=0.3640$ toward J0248$-$4048 near a pair of low-mass dwarfs at $d\approx 26$ pkpc}

\subsubsection{Absorption properties of the optically-thick gas}

The LLS at $z_{\rm abs}=0.3640$ toward J0248$-$4048 has $\tau_{\rm
  912}=2.48\pm 0.01$.  It is resolved into a minimum of six components
with $\approx 87$\% of $N(\HI)$ contained in the central component at
$d\,v_c=0$ \kms\ and $\approx 12$\% contained in the component at
$d\,v_c=-42$ \kms\ (components 3 and 2, respectively, in Table 6).
The remaining components contribute no more than 1\% to the total
$N(\HI)$, but they dominate the line width observed in the first few
Lyman lines with velocity ranging from $d\,v_c=-131$ \kms\ to
$d\,v_c=+154$ \kms\ (Figures 3 and 4).  The large $b$-value of
component 5 ($b_c\approx 69$ \kms\ and $\log\,N_c(\HI)/\cmjj\approx 14.7$)
in Table 6 is likely due to blended weak \HI\ components that are not
resolved in the COS spectra, but is necessary for explaining the
observed \lya\ and \lyb\ line profiles.  Similar to component 6 of the
LLS toward J2135$-$5316, this component is not present in higher-order
Lyman lines and the lack of resolving power has a negligible impact on
the total $N(\HI)$ measurement.  However, the best-fit $N_c(\HI)$ for
component 5 should be taken with caution.

While low-ionization
species are concentrated in the two strongest \HI\ components at
$d\,v_c=-42$ and 0 \kms\ (components 2 and 3 in Table 6), intermediate
ions exhibit a matching component structure with the first few Lyman
series lines.
In addition, the highly ionized species traced by the
\OVI\ doublet exhibit a broader line profile encompassing all
lower-ionization species, revealing a multiphase nature of the LLS.

\begin{figure}
\includegraphics[scale=0.44]{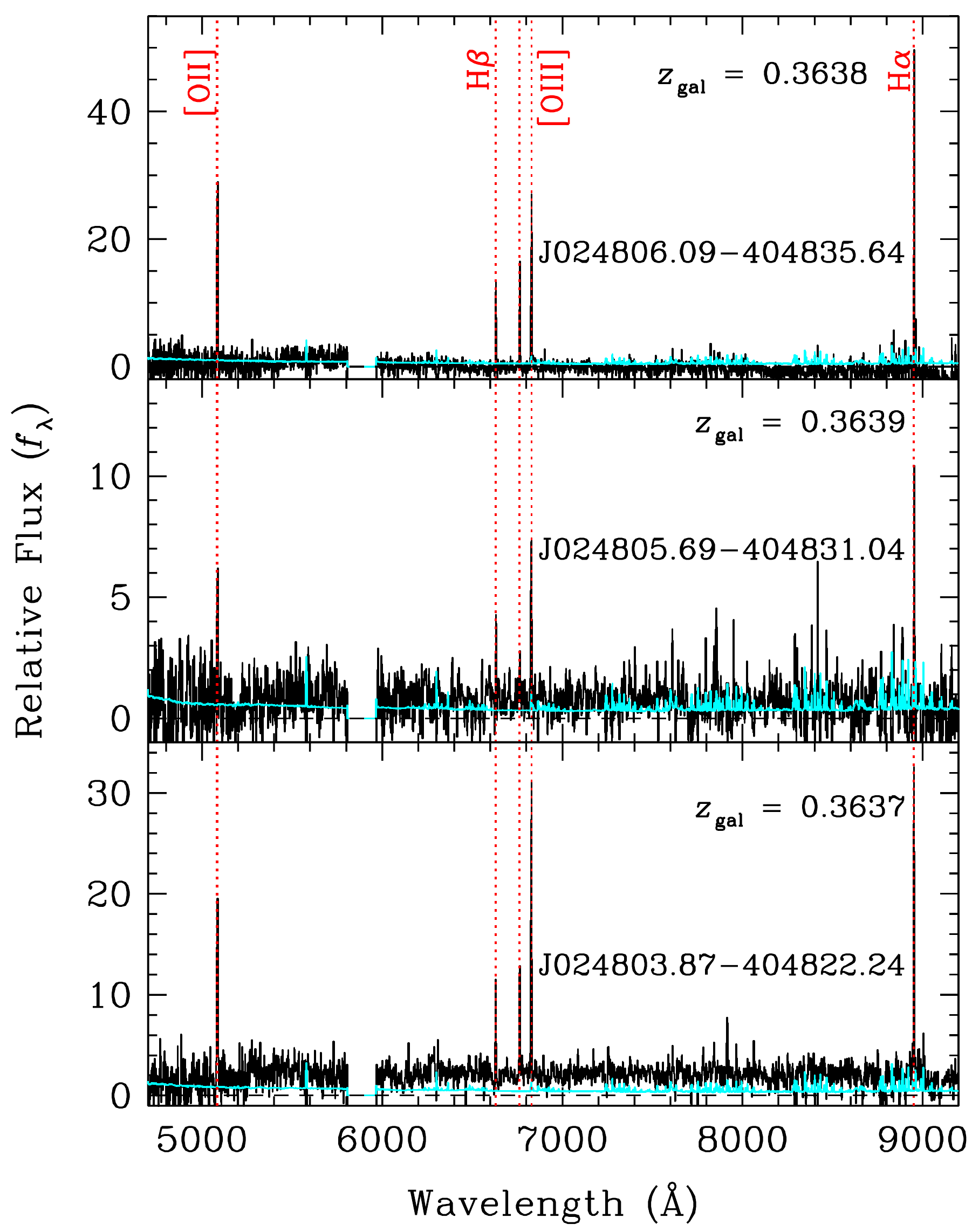}
\caption{Optical spectra of three galaxies identified in the vicinity
  of the LLS at $z_{\rm abs}=0.3640$ toward J0248$-$4048 from MUSE
  observations.  The corresponding 1-$\sigma$ error spectra are shown
  in cyan.  The spectra are ordered with increasing projected distance
  from top to bottom, at $d=15$, 37, and 150 pkpc, from the QSO
  sightline.  All three galaxies exhibit an emission-line dominated
  spectrum.  Prominent spectral features are highlighted in red dotted
  lines, marked with the corresponding line identifications.}
\label{figure:J0248gspec}
\end{figure}

\subsubsection{The galaxy environment}

The MUSE observations of this field presented in \S\ 2.4 provide a
deep view of the line-of-sight galaxy properties.  We have uncovered
67 objects in the MUSE field of view with pseudo $r$-band magnitude
ranging from $AB(r)=20.5$ to 26.9 mag in addition to the QSO.  Of the
30 objects brighter than $AB(r)=25$ mag, we are able to determine a
robust redshift for 28, reaching a survey completeness of $>90$\%.
The redshifts of spectroscopically identified galaxies in the MUSE
field around J0248$-$4048 range from $z_{\rm gal}=0.21$ to $z_{\rm
  gal}=1.46$.  Three galaxies are found in the vicinity of the LLS
with line-of-sight velocity offset $|d\,v_{\rm gal}|<70$ \kms\ (Figure
8).  The remaining spectroscopically-identified galaxies in the MUSE
field all appear at a cosmologically distinct redshift with a velocity
separation exceeding $|d\,v_{\rm gal}|=1000$ \kms\ from the LLS.  The
angular distances of the three LLS-associated galaxies are
$\theta=3.0''$, $7.8''$, and $29.7''$, corresponding to $d=15$, 37,
and 150 pkpc, respectively, at the redshift of the LLS.

The extracted MUSE spectra of the three LLS-associated galaxies are
presented in Figure 9 with increasing projected distance from top to
bottom.  All three galaxies exhibit strong nebular emission lines,
including [\OII], H$\beta$, [\OIII], and H$\alpha$, commonly seen in
star-forming regions, and they are of low mass with faint traces of
continuum emission.  Details regarding the properties of these
low-mass dwarfs are presented in Appendix B.  In summary, the galaxies
have low luminosity, $\apl\!0.03\,L_*$, and low mass,
$\log\,\mstar/\msun\apl\!8.2$, with an unobscured SFR in the range of
0.01-$0.11\,\msun\,{\rm yr}^{-1}$ and a gas-phase metallicity in their
interstellar medium (ISM) of $\approx\!10$\% solar.  The observed and
derived properties of the galaxies are shown in columns (2)
through (11) of Table 7.

While both the relative line-of-sight velocity offsets and projected
separations of the three low-mass dwarf galaxies are small with
$|d\,v_{\rm gal}|<50$ \kms\ and $d\apl 150$ pkpc to the LLS, the
observed low stellar mass also implies a low-mass host dark matter
halo.  If no massive galaxies are found in the ongoing {\it shallow
  and wide} survey beyond the MUSE field of view, then the low-mass
dwarf at $d=150$ pkpc may be too far away to be sharing a common dark
matter halo with the two galaxies at $d<40$ pkpc.  Indeed, adopting a
range of the stellar mass to dark matter halo mass ratio for low-mass
dwarfs from Miller \etal\ (2014) and Read \etal\ (2017), we infer a
dark matter halo mass of $\log\,M_h\approx 9.8-10.6$ for the galaxy at
$d=150$ pkpc and a corresponding virial radius of $r_{\rm vir}\approx
35-65$ kpc.  The LLS along the QSO sightline would be at a distance
more than $2\times r_{\rm vir}$ from the galaxy at $d=150$ pkpc.
We therefore conclude that the LLS is most likely associated with a
pair of low-mass dwarfs at $d=15$ and 37 pkpc, respectively, with an
ISM metallicity of $\approx 10$\% solar (see \S\ 5.3 below for further
discussion).

\begin{figure*}
\includegraphics[scale=0.5]{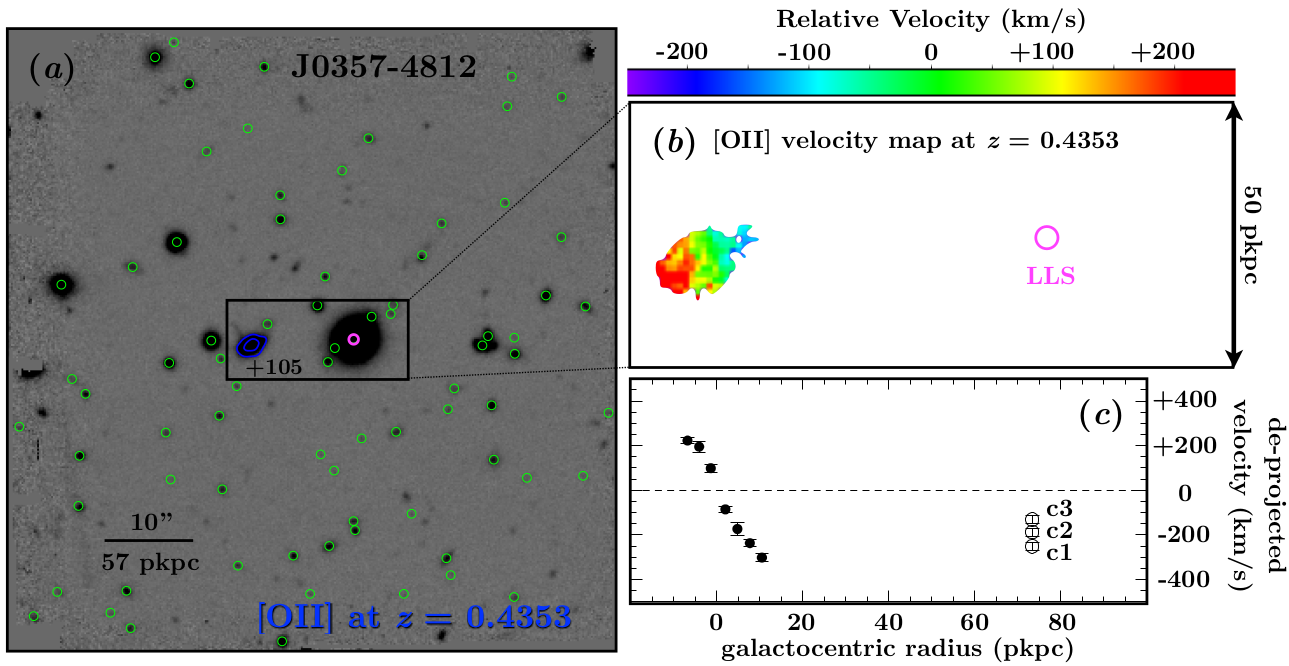}
\caption{The galaxy environment uncovered by MUSE for the LLS at
  $z_{\rm abs}=0.4353$ toward J0357$-$4812.  The pseudo $r$-band image
  from Figure 2 is reproduced in panel ({\it a}).  North is up and
  east is to the left.  Of the 77 spectroscopically identified
  galaxies in the MUSE field of view (green circles), only one galaxy
  (highlighted in blue contours) is found in the vicinity of the LLS
  at $\theta=11.9''$ (or $d=67$ pkpc) and $d\,v_{\rm gal}=+105$ \kms.  The
  [\OII] emission contours of constant surface brightness 0.25 and
  $1.25\times 10^{-17}\,{\rm erg}\,{\rm s}^{-1}\,{\rm cm}^{-2}\,{\rm
    arcsec}^{-2}$ are superimposed on top of the pseudo $r$-band
  image.  The line-of-sight velocity map of [\OII] emission is
  presented in panel ({\it b}) with zero velocity corresponding to
  $z_{\rm abs}=0.4353$.  The location of the LLS is marked by an open
  magenta circle.  The absorbing galaxy displays a clear velocity
  gradient along the long axis, supporting the presence of a rotating
  disk.  Panel ({\it c}) shows the de-projected rotation velocity
  along the disk plane as a function of galactocentric radius (solid
  points).  Here zero velocity is shifted from the LLS to the systemic
  redshift of the galaxy.  For comparison, the three strongest
  \HI\ absorbing components (c1, c2, and c3 from Table 6) of the LLS
  along the QSO sightline are also included in the plot, under the
  assumption that the gas is moving along the plane extended from the
  inclined optical disk.  The de-projected velocities of the three
  components are consistent with a general rotation motion with the
  galaxy (see text for further discussion).}
\label{figure:J0357gmap}
\end{figure*}

\subsection{The LLS at $z_{\rm abs}=0.4353$ toward J0357$-$4812 near a rotating disk galaxy at $d=67$ pkpc}

\subsubsection{Absorption properties of the optically-thick gas}

The LLS at $z_{\rm abs}=0.4353$ toward J0357$-$4812 has $\tau_{\rm
  912}=0.99\pm 0.01$.  It is resolved into a minimum of six components
with $\apg 99$\% of $N(\HI)$ contained in the central three components
at $|d\,v_c|<40$ \kms\ (components 1 through 3 in Table 6).  The
remaining components contribute negligibly to the total $N(\HI)$, but
are necessary for explaining the observed line profiles of the first
few Lyman lines with velocity ranging from $d\,v_c=86$ \kms\ to
$d\,v_c=+252$ \kms\ (Figures 3 and 4).

Figure 4 shows that the component at $d\,v_c=+31.4$ \kms\ (component 3
of this LLS in Table 6) exhibits the strongest metal absorption, while
containing only $\approx 8$\% of the total $N(\HI)$.  Specifically,
components 2 and 3 exhibit a column density ratio of 10:1 in $N(\HI)$,
but 4:9 in $N(\MgII)$.  This differential absorption strength between
\HI\ and associated ionic transitions of components 2 and 3 applies to
all ions observed from low- to high-ionization species, strongly
implying a difference in the underlying gas metallicity (e.g.,
Prochter \etal\ 2010; Zahedy \etal\ 2019).  Similar to the LLSs toward
J2308$-$5258 and J0248$-$4048, the highly ionized species traced by
the \OVI\ doublet exhibit a broad line profile encompassing all
lower-ionization species, indicating a multiphase nature of the LLS.

\begin{figure}
\includegraphics[scale=0.35]{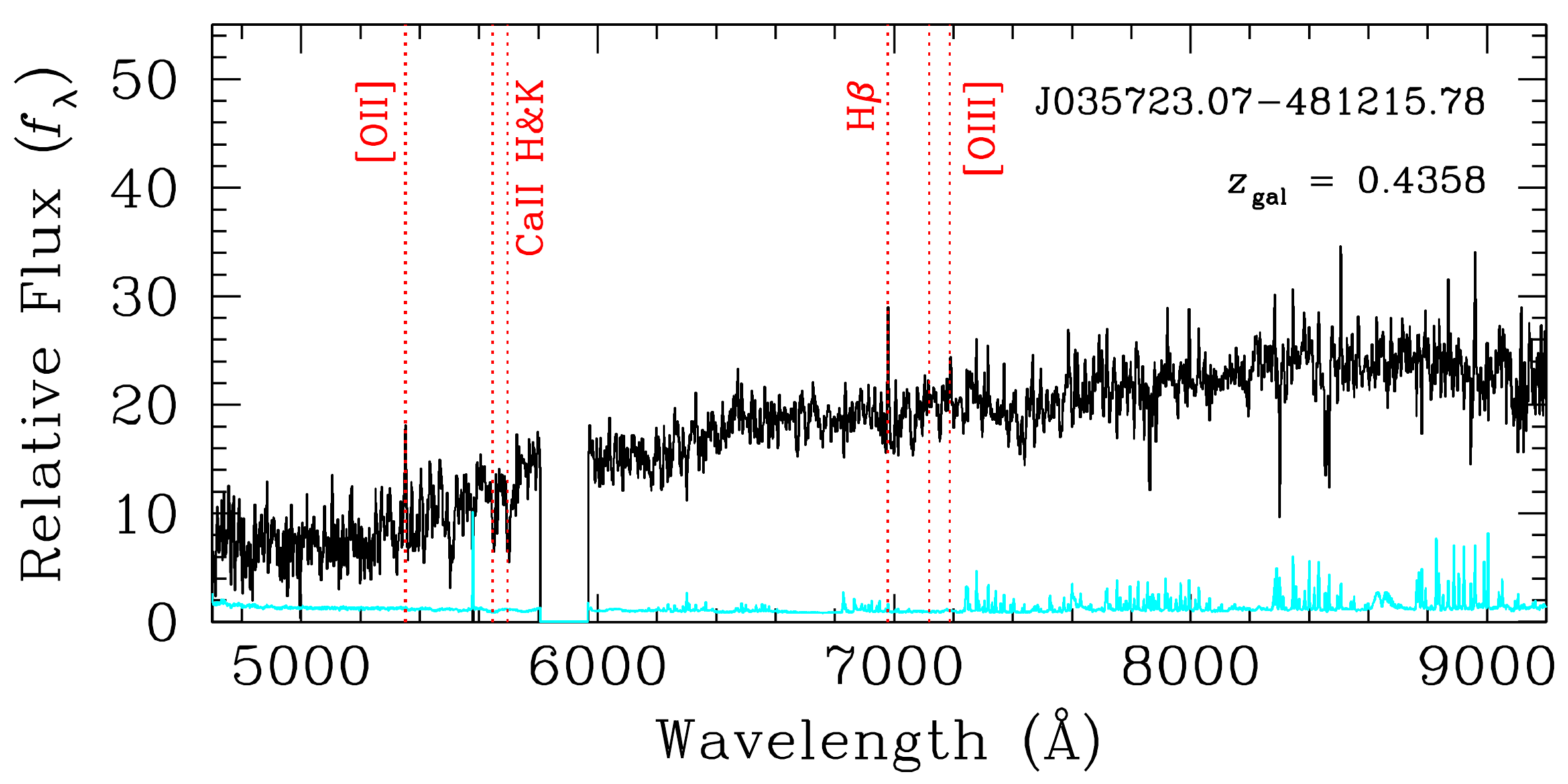}
\caption{Optical spectrum of the galaxy identified in the vicinity of
  the LLS at $z_{\rm abs}=0.4353$ toward J0357$-$4812 from MUSE
  observations.  The corresponding 1-$\sigma$ error spectrum is shown
  in cyan.  The galaxy is bright and the spectrum exhibits prominent
  absorption features along with weak [\OII] and H$\beta$ emission
  lines (highlighted in red dotted lines with corresponding line
  identifications).  In addition to the \CaII\ absorption doublet, the
  spectrum also exhibits strong Balmer absorption lines, indicating
  the presence of mixed stellar populations.}
\label{figure:J0357gspec}
\end{figure}

\subsubsection{The galaxy environment}

The MUSE observations of this field presented in \S\ 2.4 have
uncovered 129 objects in the MUSE field of view with pseudo $r$-band
magnitude ranging from $AB(r)=20.5$ to 27 mag in addition to the QSO.
Of the 64 objects brighter than $AB(r)=25$ mag, we are able to
determine a robust redshift for 51, reaching a survey completeness of
$80$\%.  The redshifts of spectroscopically identified galaxies in
this field range from $z_{\rm gal}=0.20$ to $z_{\rm gal}=3.668$.
Similar to the LLS toward J2308$-$5258 (see \S\ 4.3), only one galaxy
is found in the vicinity of the LLS at line-of-sight velocity offset
$d\,v_{\rm gal}=+105$ \kms\ and angular distance $\theta=11.9''$,
corresponding to $d=67$ pkpc at the redshift of the LLS (Figure 10).
Different from the configuration between the QSO probe and the LLS
absorbing galaxy in J2308$-$5258, the QSO here probes the diffuse gas
at $\approx 27^\circ$ from the major axis of the inclined disk (see
below).  The remaining spectroscopically-identified galaxies in the
MUSE field are all at a cosmologically distinct redshift with a
velocity separation exceeding $|d\,v_{\rm gal}|=1000$ \kms\ from the
LLS.  The observed $r$-band magnitude of the LLS-absorbing galaxy is
$AB(r)=20.7$, while objects without a robust redshift measurement are
all fainter than $AB(r)=23.3$ mag.  Any additional galaxy in the
vicinity of the LLS will have to be more than 10 times fainter with
intrinsic luminosity $<0.14\,L_*$.

The optical spectrum of the absorbing galaxy at $d=67$ pkpc is
presented in Figure 11, which is characterized by strong absorption
features, together with weak [\OII] and H$\beta$ emission lines.  In
particular, it exhibits prominent \CaII\ absorption along with strong
Balmer absorption lines.  Details regarding the properties of this
galaxy are presented in Appendix C.  In summary, the galaxy is
luminous with an unobscured SFR of $\approx 0.28\,\msun\,{\rm
  yr}^{-1}$ based on the total integrated [\OII] line flux and the
star formation calibrator of Kewley \etal\ (2004).  The observed
$r$-band magnitude, $AB(r)=20.7$, and rest-frame $g-r$ color translate
to an intrinsic $r$-band absolute magnitude of $M_r=-21.7$
(corresponding to $1.4\,L_*$) based on an early-type disk galaxy
template and an underlying stellar mass of $\log\,\mstar/\msun=10.4$.
The observed and derived properties of the galaxy are presented in
columns (2) through (11) of Table 7.

The line-of-sight velocity map presented in panel ({\it b}) of Figure
10 shows that the absorbing galaxy displays a clear velocity gradient
along the long axis, from $\approx +250$ \kms\ on the southeast
(lower-left) side to $\approx -110$ \kms\ on the northwest
(upper-right) side of the galaxy with the zero velocity corresponding
to the systemic redshift, $z=0.4353$, of the strongest component (c2
in Table 6) in the LLS at $d=67$ pkpc to the west.  The de-projected
rotation velocity along the disk plane as a function of galactocentric
radius is presented in panel ({\it c}) of Figure 10 (solid points)
with zero velocity corresponding to the systemic redshift of the
galaxy (see Appendix C for details).  Adopting $d\,v_{\rm
  deproj}\approx +225$ \kms\ to be the maximum rotation velocity would
imply an underlying dark matter halo mass of $M_h\approx
10^{12}\,\msun$ and a stellar mass to dark matter halo mass ratio of
$\mstar/M_h\approx 2.5$\%.  The three \HI\ absorbing components
containing $>99$\% of the total $N(\HI)$ (c1, c2, and c3 from Table 6)
of the LLS are found to be moving along the same direction as the
optical disk at a velocity, ranging from $d\,v_{\rm deproj}({\rm
  LLS})=-130$ to $-250$ \kms\ (open circles in panel {\it c} of Figure
10).  The observed co-rotation between the gas and the galaxy suggests
that the LLS originates in a rotating halo or an extended disk plane
(e.g., Steidel \etal\ 2002; Chen \etal\ 2005; Diamond-Stanic
\etal\ 2016; Martin \etal\ 2019) with contributions from either
extraplanar gas and/or high velocity clouds (e.g., Heald \etal\ 2011).
Interestingly, component 3 with the largest $N(\MgII)$ (and likely
higher gas metallicity; see discussion in \S\ 5.2) also exhibits the
largest lag in $|d\,v_{\rm deproj}|$ (see Figure 4).

\section{Discussion}
\label{section:sample}

The LLS survey in the CUBS QSO sample has uncovered five new LLSs with
$\tau_{912}\apg 1$ at $z_{\rm abs}=0.3640-0.6226$ and five pLLS with
$\tau_{912}=0.2-1$ at $z_{\rm abs}=0.2603-0.9372$.  While the number
of newly identified LLSs is small, these represent a uniform and
unbiased sample of optically-thick absorbers identified along random
QSO sightlines with $\log\,N(\HI)/\cmjj\apg 16.5$.  Common features of
the newly identified LLSs include: (1) multi-component kinematic
profiles, indicating a clumpy medium, and (2) a simultaneous presence
of ionic species in multiple ionization states, suggesting a
multiphase nature of the gas.  These features have important
implications for studying the ionization state and metallicity of the
gas.

Deep galaxy survey data in the LLS fields have revealed a diverse
range of galaxy environments around these optically-thick absorbers,
from a massive quiescent halo, a low-mass dwarf galaxy pair, a
co-rotating gaseous halo/disk, to possibly intragroup medium of a
galaxy group.  A joint analysis, combining absorption-line
studies of newly discovered LLSs and the associated galaxy survey data
from the CUBS program, has offered important new insights into the
connection between optically-thick gas and galaxies, as well as the
physical origin of metal-line absorbers.  In this section, we discuss
the implications of our sample for the incidence of LLSs at $z<1$ and
the physical properties of these absorbers, and the galactic
environment uncovered from our galaxy survey.

\subsection{Incidence of Lyman Limit Systems at $z<1$}

Because of their optically thick nature, the rate of incidence of LLSs
is an important quantity for constraining the mean free path of
ionizing photons and for computing the IGM photo-ionization rate.  It
also places constraints on the covering fraction of optically-thick
gas connected to infalling clouds and/or outflows in galaxy halos
(e.g., Hafen \etal\ 2017).  Previous surveys have yielded a number
density of LLSs per unit redshift per sightline in the range of
$n(z)\approx 0.3-0.7$ at $z<1$ (e.g., Songaila \& Cowie 2010 for a
compilation of early results with updates from Ribaudo \etal\ 2011;
Shull \etal\ 2017).  The large uncertainty underscores the challenge
in conducting LLS surveys.  First, surveys for LLSs at low redshifts
rely on UV spectroscopy from space and therefore require a sample of
UV-bright QSOs at $z_{\rm QSO}\apg 1$ as the background probes.
Secondly, these systems are rare and when a LLS occurs, it also
attenuates the background QSO light and further reduces the survey
path and sensitivity for uncovering additional Lyman limit features at
lower redshift.  This is seen in two of the CUBS QSO sightlines in
Table 5, J0111$-$0316 and J2135$-$5316, which exhibit the two
strongest LLSs found in our survey and consequently $\Delta\,z_{\rm
  LL}$ is substantially reduced for those two sightlines.

\begin{figure}
\includegraphics[scale=0.25,angle=0]{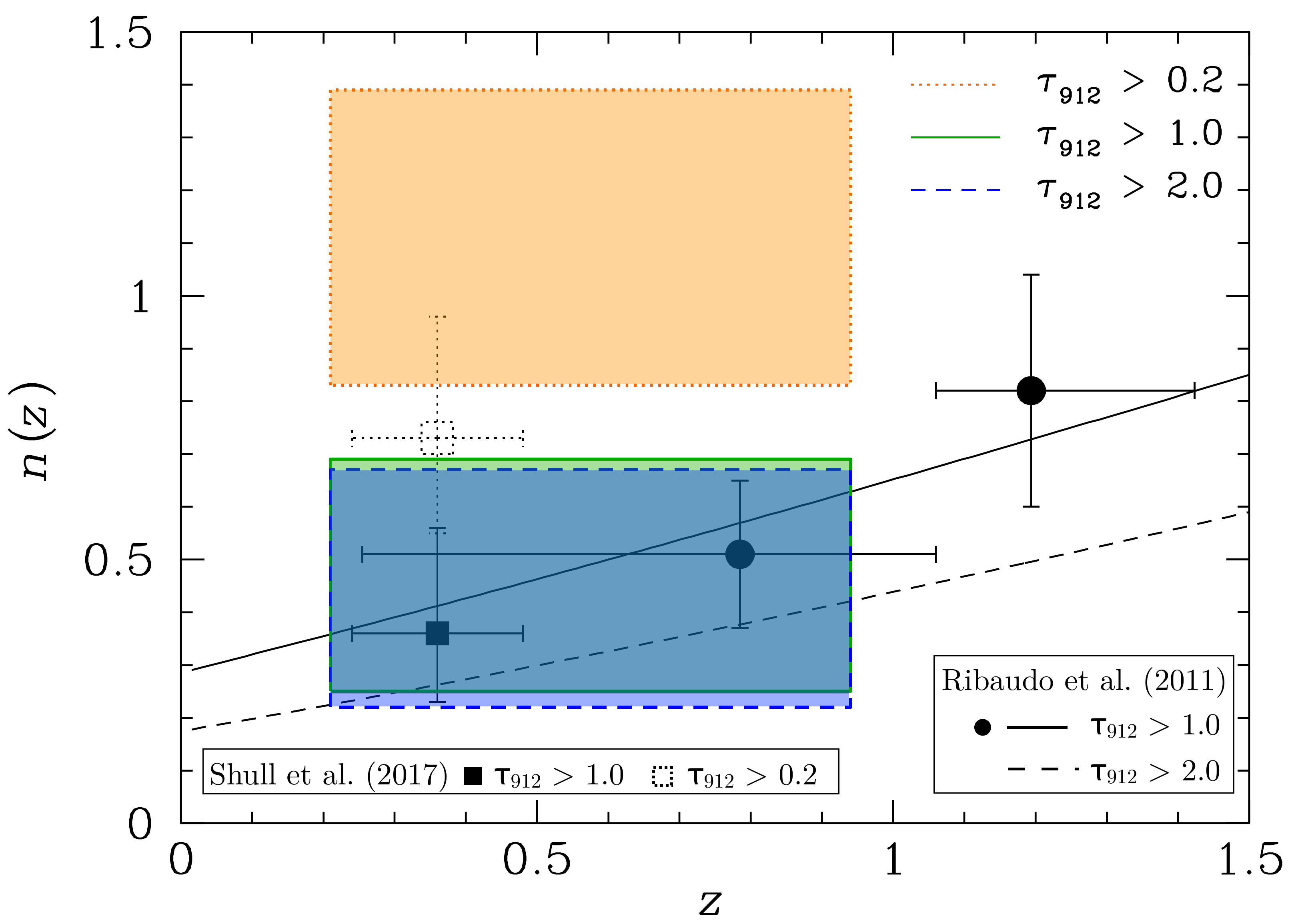}
\caption{The rate of incidence of optically-thick \HI\ absorbers at
  $z\apl 1$ from the CUBS QSO sample.  The measurements are made for
  three different $\tau_{912}$ thresholds, from $\tau_{912}>2$ (blue,
  dashed rectangle), $\tau_{912}>1$ (green, solid rectangle), to
  $\tau_{912}>0.2$ (orange, dotted rectangle).  The height of each
  rectangle represents the 68\% confidence interval estimated from a
  bootstrap resampling exercise described in the text, and the width
  represents the redshift interval surveyed.  Previous measurements
  and associated 1-$\sigma$ uncertainties from Ribaudo \etal\ (2011)
  are shown in solid points, along with their best-fit redshift
  evolution models for $\tau_{912}>1$ (solid curve) and for
  $\tau_{912}>2$ (dashed curve).  The measurements and 1-$\sigma$ errors
  from Shull \etal\ (2017) are shown in squares.}
\label{figure:nz}
\end{figure}

The consequence of attenuated background QSO light when a LLS occurs
has important implications for the sample definition of QSOs employed
for an unbiased CGM survey.  UV-bright QSOs are biased against finding
LLSs which, when unaccounted for, would lead to an underestimate of
the incidence of these optically thick systems (e.g., Tytler 1982).
Previous LLS surveys relying on archival sightlines to maximize the
redshift survey pathlength for these relatively rare systems have
attempted to minimize such a bias by excluding QSO sightlines that are
known to be biased against LLSs (e.g., Ribaudo \etal\ 2011).  However,
the sample is inherently heterogeneous and likely contains implicit
bias.  A recent study of the GALEX FUV-NUV colors of a sample of over
9000 NUV-selected QSOs (see the illustration in the top panel of
Figure 1) indeed suggests that the number density of LLSs at $z<1$ may
have been underestimated (Deharveng \etal\ 2019).

The CUBS QSO sample is selected to be at $z_{\rm QSO}\approx 0.8$--1.4
and bright in the GALEX NUV channel (1770--2730 \AA).  As outlined in
\S\ 2, these selection criteria are motivated by a key scientific
goal of studying the inner CGM ($d\apl 100 $ pkpc) at $z\apl 1$ where
a high incidence of optically-thick absorbers is expected.  The QSO
sample is therefore also optimized for an unbiased search of LLS at
$z<1$.  The summary in Table 5 shows that the entire CUBS QSO sample
offers a total survey pathlength of $\Delta\,z_{\rm LL}=9.31$ for LLSs
at $z=0.21$--0.94.  To estimate the number density of low-redshift LLSs
and the associated uncertainties, we perform a bootstrap resampling
routine.  Specifically, we randomly resample the parent CUBS QSO
sightlines with replacement to generate a new QSO sample containing
the same number of QSOs.  For each Lyman limit decrement in the new
QSO sample, a new $\tau_{912}$ is generated from randomly sampling the
Gaussian error around the measured $\tau_{912}$ reported in Table 5.
This exercise is repeated 1000 times to produce a distribution of newly
estimated number density $n(z)$ of the LLSs.  The resulting 16 and 84
percentiles mark the 68\% confidence interval of the best estimated
$n(z)$.

Our survey yields a number density estimate of
$n(z)=0.43_{-0.18}^{+0.26}$ for systems with $\tau_{912}>1$.  This is
consistent with expectations from previous archival studies by Ribaudo
\etal\ (2011) and Shull \etal\ (2017).  It is, however, evident from
Table 5 that most of the newly discovered LLSs in the CUBS QSO sample
exhibit $\tau_{912}$ significantly greater than one.  We repeat the
number density measurement for two different thresholds with
$\tau_{912}>2$ and $\tau_{912}>0.2$.  The results are presented in
Figure 12, along with measurements from previous archival studies for
comparison.  For $\tau_{912}>2$, we find $n(z)=0.43_{-0.21}^{+0.24}$
in comparison to $n(z)=0.37\pm 0.10$ from Ribaudo \etal\ (2011).  For
$\tau_{912}>0.2$, we find $n(z)=1.08_{-0.25}^{+0.31}$ in comparison to
$n(z)=0.73_{-0.18}^{+0.23}$ from Shull \etal\ (2017).  Our best
estimated $n(z)$ for these strong \HI\ absorbers are higher than
previously reported values, but the uncertainties are large due to the
small QSO sample.

\subsection{The Origin of Lyman Limit Systems}

We have shown in \S\ 3.2 that in order to fully characterize the Lyman
series associated with each LLS, it is necessary to incorporate a
multi-component structure for the absorber.  Specifically, while the
strongest component explains the highest-order Lyman series lines all
the way to the Lyman discontinuity, additional components are needed
to explain the first few Lyman lines with the largest oscillator
strengths (see also Chen \etal\ 2018; Zahedy \etal\ 2019).  This
multi-component structure is also supported by the associated ionic
transitions.  In particular, nearly all \HI\ components of
$\log\,N(\HI)/\cmjj>16.8$ exhibit associated \MgII\ components that
are fully resolved in the ground-based optical echelle spectra (Figure
4).  While COS does not fully resolve the component structure of
additional FUV lines such as \CII, \CIII, \NII, and \SiII, there is a
broad agreement in the overall line shape (including both line width
and centroid) of these FUV transitions with the relative component
strength displayed in the resolved \MgII\ components.  The agreement
in the absorption profiles between low- and intermediate-ionization
transitions supports a scenario that these ions are linked
kinematically and share a common velocity field.

\begin{figure*}
\includegraphics[scale=0.325,angle=0]{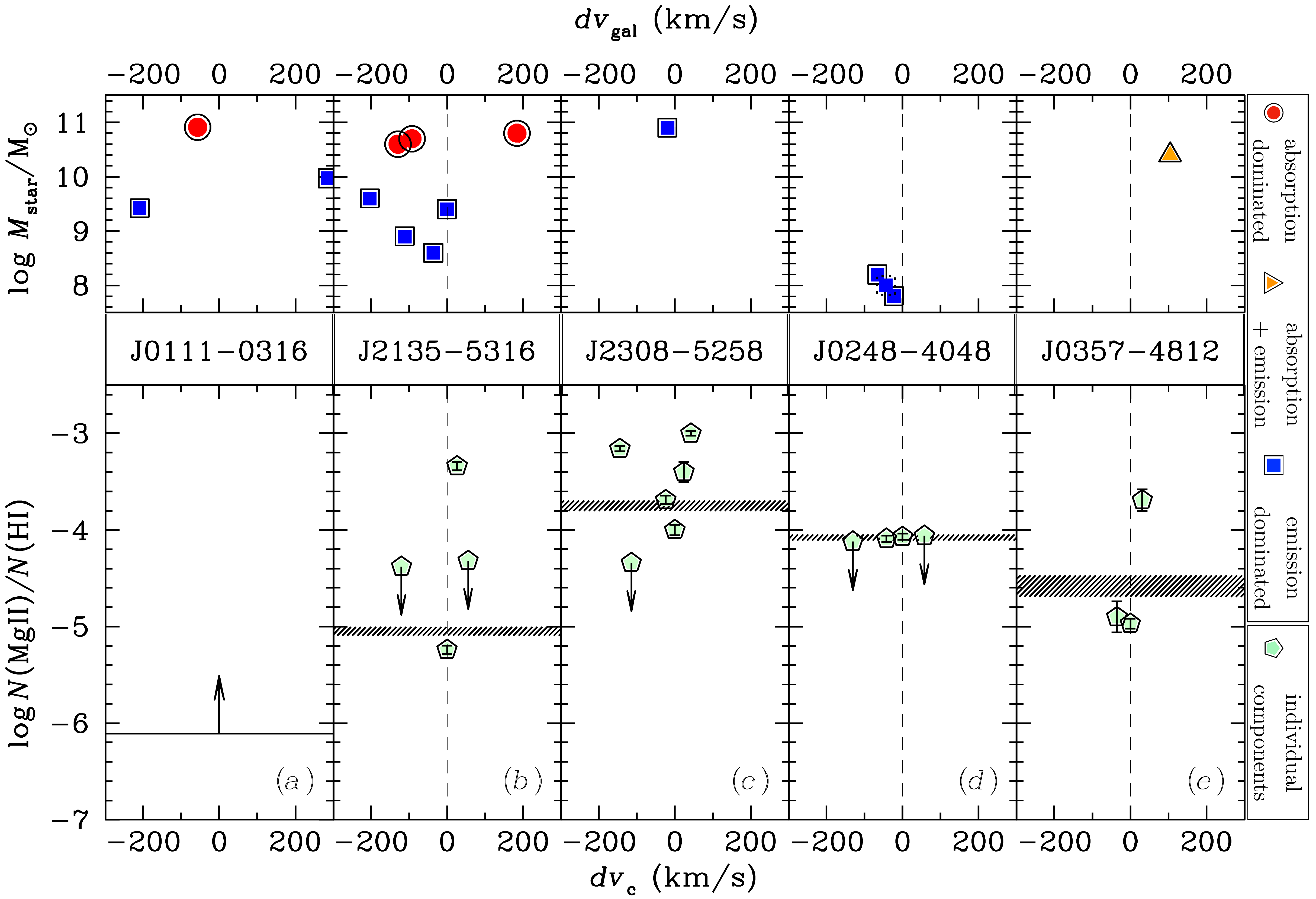}
\caption{{\it Top}: Stellar mass versus velocity distribution of
  galaxies identified at $d<300$ pkpc and $|d\,v_{\rm gal}|<300$
  \kms\ from the LLS (see Table 7).  Blue squares represent
  emission-line dominated star-forming galaxies; red circles represent
  absorption-line (such as \CaII\,H\&K) dominated galaxies; and orange
  triangles represent galaxies displaying both emission and absorption
  in their optical spectra.  The square with a dotted outline
  represents the galaxy at $d = 15$ pkpc toward J0248$-$4048, because
  its mass is inferred from the estimated gas-phase metallicity and
  the mass-metallicity relation of Berg \etal\ (2012).  {\it Bottom}:
  Observed $N(\MgII)/N(\HI)$ ratios of five new LLSs from the CUBS
  sample versus the line of sight velocity offset from the strongest
  \HI\ component ($d\,v_c$ from Table 6).  The fields are ordered with
  decreasing $N(\HI)$ from left to right.  Pentagons are for
  individual components of $\log\,N(\HI)/\cmjj>15$, while the
  horizontal bar in each panel represents the integrated column
  density ratio and its associated uncertainty over all components for
  each LLS.  Large scatter between individual components is present in
  three of the four strong \HI\ absorbers for which the component
  structure is resolved for both \MgII\ and \HI\ (panels {\it b}, {\it
    c}, and {\it e}), and absorbing gas with high $N(\MgII)/N(\HI)$
  would be missed in integrated column density ratios. }
\label{figure:nz}
\end{figure*}

We also note that \OVI\ absorption is observed in all but the DLA at
$z_{\rm abs}=0.5762$ toward J0111$-$0316.  Three of the four
\OVI-bearing LLSs, toward J0248$-$4048, J0357$-$4812, and
J2308$-$5258, display a broad and relatively shallow \OVI\ absorption
profile.  While the component structures of the \OVI\ lines are not
resolved, their kinematic profiles share a similar asymmetry with the
associated \CIII\ absorption line.  Such agreement suggests a
multiphase nature of these LLSs (see also, Zahedy \etal\ 2019).
A notable exception is the
LLS at $z_{\rm abs}=0.6226$ toward J2135$-$5316, which displays a
single strong \OVI\ component at $d\,v_c=-120$ \kms\ with
corresponding \CIII\ and a relatively weak \HI\ component of
$\log\,N_c(\HI)/\cmjj=15.4$ and $b_c\approx 27$ \kms, but no other
low-ionization transitions.  The co-presence of \CIII\ and \OVI\ with
consistent kinematic profiles with \HI\ suggests that the gas is
predominantly photo-ionized (e.g., Tripp \etal\ 2008; Thom \& Chen
2008; Stern \etal\ 2018).  However, the absence of \OVI\ around
dominant \HI\ components at $|d\,v_c|<100$ \kms\ is in stark contrast
to the other three LLSs toward J0248$-$4048, J0357$-$4812, and
J2308$-$5258.

Comparisons of heavy ions to neutral hydrogen column density ratios
between individual components have also revealed a chemical
inhomogeneity within a single LLS.  Figure 13 shows the observed
$N(\MgII)/N(\HI)$ ratios both for individual components (pentagons)
and for each LLS as a whole (horizontal bars).  Only components with
$\log\,N(\HI)/\cmjj>15$ are included, because the sensitivity limit
for $N(\MgII)$ allowed by the optical echelle spectra become
unconstraining for weaker \HI\ components.  In at least three of the
five new LLSs in the CUBS sample, large differential $N(\MgII)/N(\HI)$
is directly seen between different components.  For example, the
strongest \MgII\ component of the LLS at $z_{\rm abs}=0.4353$ toward
J0357$-$4812 ($c3$ in Table 6) occurs at $d\,v_c=+31.4$ \kms\ and
contains 10 times less \HI\ than the strongest \HI\ component ($c2$ in
Table 6) at $d\,v_c=0$ \kms.  Consequently, the two components exhibit
$N(\MgII)/N(\HI)$ ratios that differ by more than a factor of 10
(panel {\it e} of Figure 13).

Although inferring gas metallicity from the observed $N(\MgII)/N(\HI)$
requires knowledge of the ionization state of the gas (which is
presented in a separate paper by Zahedy \etal\ 2020, in preparation),
we note that under photoionization equilibrium the observed
differences in $N_c(\MgII)/N_c(\HI)$ could serve as a rough guide for
inferring the underlying gas metallicities for a limited gas density
range under photo-ionization equilibrium.  This is understood based on
the expectation that the abundance ratio of total Mg/H is related to
$N(\MgII)/N(\HI)$ according to $({\rm Mg}/{\rm H})\times(f_{{\rm
    Mg}^+}/\,f_{{\rm H}^0})={\rm Mg}^+/{\rm H}^0$.  For a typical gas
density of diffuse halo gas in the range $n_{\rm H}\approx
0.01$--$0.001\,{\rm cm}^{-3}$ (corresponding to an ionization
parameter in the range of $\log\,U\approx -3$ -- $-2$ at $z<1$,
assuming ionization equilibrium and no dust depletion), the ratio of
ion fractions, $f_{{\rm Mg}^+}/\,f_{{\rm H}^0}$, does not change by
more than a factor of two (e.g., Chen \& Tinker 2008; Wotta
\etal\ 2016).  The large variation by more than a factor of 10 in the
observed $N(\MgII)/N(\HI)$ between individual components therefore
suggests large differential chemical enrichment levels in individual
LLSs (see also Prochter \etal\ 2010; Zahedy \etal\ 2019).

In summary, newly uncovered LLSs in the CUBS QSO sample exhibit a
multi-component absorption structure with differential chemical
enrichment levels as well as ionization states.  The gas is complex in
terms of its kinematics and ionization state, and inhomogeneous in
chemical enrichment.  These properties have important implications
both for the physical origins of these optically-thick absorbers and
for ionization and metallicity measurements that are based on
integrated column density ratios over all components.  They also
underscore the scientific value of high-resolution absorption spectra
in resolving the clumpy, multiphase gas (e.g., Zahedy \etal\ 2019;
Rudie \etal\ 2019).

\subsection{Galaxy Environment of Lyman limit Systems}

\begin{figure}
\includegraphics[scale=0.45,angle=0]{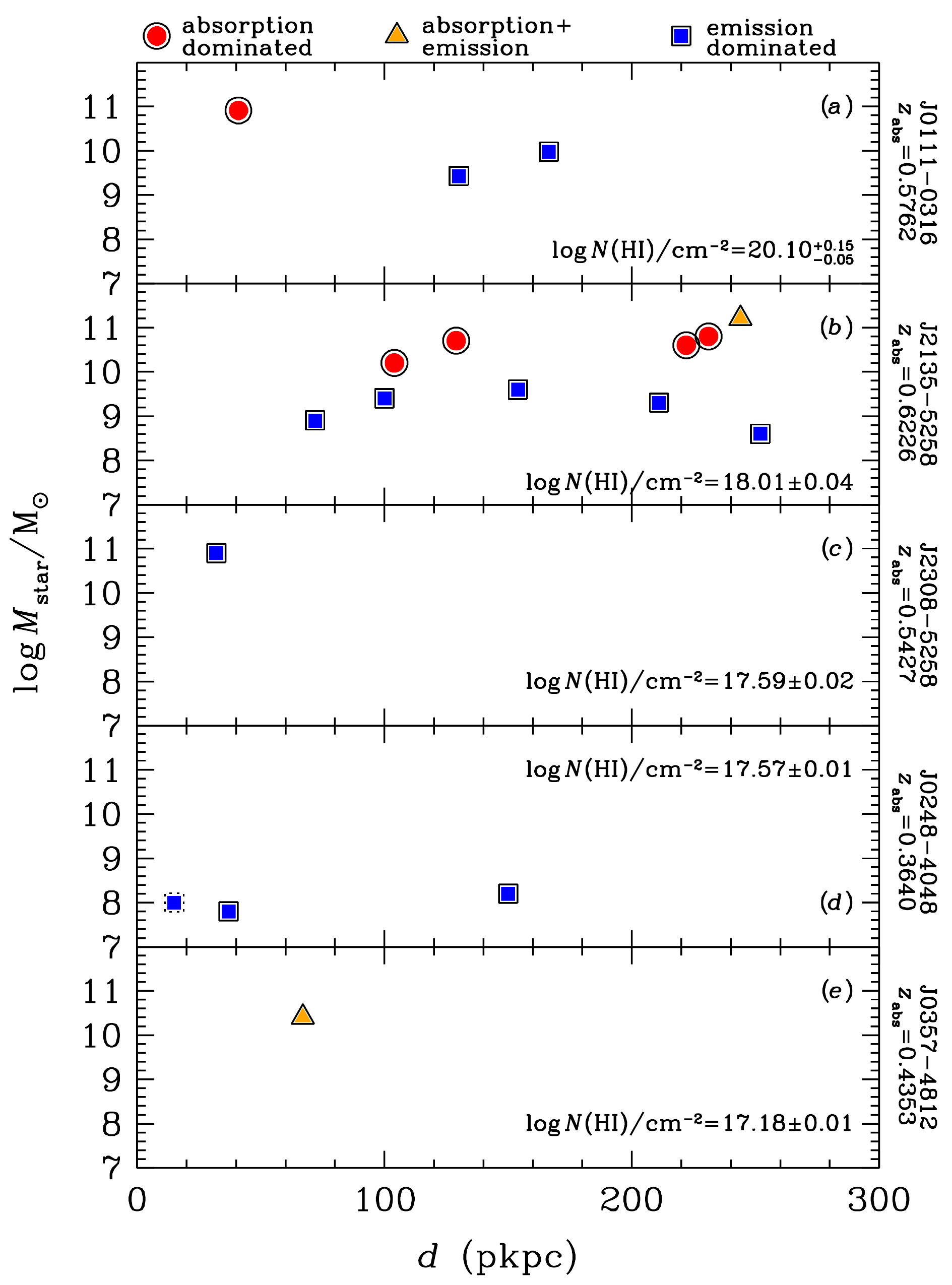}
\caption{Summary of the galaxy environment uncovered from follow-up
  deep galaxy surveys of the five new optically-thick absorbers in the
  CUBS sample.  The panels display \mstar\ versus $d$ for galaxies
  found in the vicinities of the LLSs along the sightlines toward
  J0111$-$0316 (panel {\it a}; from Boettcher et al. 2020),
  J2135$-$5316 (panel {\it b}), J2308$-$5258 (panel {\it c}),
  J0248$-$4048 (panel {\it d}), and J0357$-$4812 (panel {\it e}).  The
  fields are ordered with decreasing $N(\HI)$ from top to down.  Blue
  squares represent emission-line dominated star-forming galaxies; red
  circles represent absorption-line (such as \CaII\,H\&K) dominated
  galaxies; and orange triangles represent galaxies displaying both
  emission and absorption in their optical spectra.  The galaxy at $d
  = 15$ pkpc toward J0248$-$4048 is included for comparison (shown in
  dotted outline in panel {\it d}), but note that, as discussed in
  \S\ 4.4, the mass is inferred from the estimated gas-phase
  metallicity and the mass-metallicity relation of Berg \etal\ (2012).
  MUSE data enable a galaxy survey to faint magnitudes, reaching 100\%
  completeness at $AB(r)\le 23$ mag and $\apg 75$\% at $AB(r)\le 25$
  mag, in the field of view that corresponds roughly to an area of
  $\approx 300$ pkpc in radius at the redshifts of these absorbers.
  The absence of additional galaxies in the vicinities of the two LLSs
  toward J0357$-$4812 and J2308$-$5258 therefore indicates that the
  LLSs and the associated galaxies (both are super-$L_*$) reside in a
  relatively isolated environment.}
\label{figure:nz}
\end{figure}

In addition to complex gas kinematics, ionization states, and chemical
enrichment histories of the gas revealed in the QSO absorption
spectra, the accompanying galaxy survey has also revealed a diverse
range of galaxy environments around the new LLSs in the CUBS sample.
To summarize the galaxy environment of the LLSs, we show in the top
panels of Figure 13 the stellar mass versus velocity distribution of
galaxies identified at $d<300$ pkpc and $|d\,v_{\rm gal}|<300$
\kms\ from the LLS.  In addition, we present the \mstar\ versus $d$
distribution in Figure 14.  It shows the properties of galaxies
identified at $d<300$ pkpc of the LLSs broadly classified into three
different spectral types: evolved galaxies with an absorption-line
dominated spectrum (circles), intermediate galaxies displaying both
nebular emission and stellar absorption features (triangles), and
star-forming galaxies displaying predominantly nebular emission lines
(squares).  As described in \S\ 4, the MUSE data enable a highly
complete galaxy survey to faint magnitudes, reaching 100\%
completeness at $AB(r)\le 23$ mag and $\apg 75$\% at $AB(r)\le 25$
mag, in the field of view that corresponds roughly to an area of
$\approx 300$ pkpc in radius at the redshifts of these absorbers.  The
absence of additional galaxies in the vicinities of the two LLSs
toward J2308$-$5258 (panel {\it c}) and J0357$-$4812 (panel {\it e})
is therefore particularly interesting.  At the redshifts of these two
LLSs, our galaxy survey completeness ensures that any additional
galaxy in the vicinity of the LLS would be faint with intrinsic
luminosity $\apl 0.1\,L_*$.  It indicates that the LLSs and the
associated absorbing galaxies (both super-$L_*$ and massive) reside in
a relatively isolated environment.  Galaxies identified at $d<300$
pkpc of the ${\rm H}_2$-bearing DLA are included for comparison.  As
described in Boettcher \etal\ (2020), six additional galaxies are
found at $|d\,v_c|<300$ \kms\ at $d=300-600$ pkpc from the ongoing
deep- and wide-field galaxy survey on the Magellan Telescopes.
However, within the loose group of nine galaxies, the evolved galaxy
at $d=42$ pkpc is the most massive member, contributing $\approx 40$\%
of the total stellar mass of the group.

Similarly, the LLS at $z_{\rm abs}=0.6226$ toward J2135$-$5316 is
found associated with a massive galaxy group of dynamical mass $M_{\rm
  dyn}\approx 1.1\times 10^{13}\,\msun$ (comparable in mass to halos
hosting luminous red galaxies; Padmanabhan \etal\ 2007) at $d_{\rm
  group}=177$ pkpc.  The massive group environment provides a
potential physical explanation for the lack of \OVI\ surrounding the
strong \HI\ components (see Figure 4).  Because the gas temperature in
a massive group is expected to be high, $T\sim 6\times 10^6$ K, heavy
elements, including oxygen, are expected to be ionized to higher
states (e.g., Oppenheimer \etal\ 2016).  Indeed, Wijers \etal\ (2020)
predict covering fractions of only $\approx 25$\% for \OVI\ gas of
$\log\,N(\OVI)/\cmjj>13.5$ at $d\approx 200$ pkpc from a galaxy with
$\log\,\mstar/\msun>11$.  At the same time, the likely photo-ionized
\OVI\ component at $d\,v_c=-120$ \kms\ with $\log\,N(\HI)/\cmjj=15.4$
may be attributed to halo clouds associated with the nearby low-mass
satellite of $\log\,\mstar/\msun=8.9$ at $d=72$ pkpc and
$\Delta\,v=-111$ \kms.  However, the dominant \HI\ component of this
LLS contains the second highest $N(\HI)$ uncovered in our survey with
$\log\,N(\HI)/\cmjj=18$ and only relatively weak \MgII\ absorption.
The low $N(\MgII)/N(\HI)$ ratio (panel {\it b} of Figure 13) implies a
relatively low gas metallicity.  It is therefore likely that the LLS
originate in dense gaseous streams accreted from the IGM (e.g., Hafen
\etal\ 2017).

In contrast to the two strongest \HI\ absorbers uncovered in our
survey, the LLS at $z_{\rm abs}=0.3640$ toward J0248$-$4048 is found
to be associated with a pair of low-mass dwarfs of $\approx
0.01$--$0.03\,L_*$ ($\log\,\mstar/\msun\approx 8$) and ISM metallicity of
$\approx 10$\% solar at $d=15$ and 37 pkpc.  The galaxy at $d=15$ pkpc
is the closest absorbing galaxy uncovered in our survey.
However, the total \HI\ column density is
merely $\log\,N(\HI)/\cmjj=17.6$.  This low $N(\HI)$ at a small distance from a low-mass dwarf is similar to what is observed in local dwarf irregular galaxies (e.g., Hunter \etal\ 2012; Johnson  \etal\ 2017).
The low-mass nature (and presumably shallow
gravitational potential well) in the presence of the UV background
makes it difficult to maintain a large amount of neutral gas (e.g.,
Johnson \etal\ 2017).  The two dominant \HI\ components at
$d\,v_c=-42$ and $d\,v_c=0$ \kms\ ($c2$ and $c3$) of this LLS exhibit
comparably modest $N_c(\MgII)/N_c(\HI)$ ratios (panel {\it d} of
Figure 13), suggesting a chemical enrichment level comparable to the
10\% solar metallicity (e.g., Wotta \etal\ 2016) inferred for the ISM
of the two star-forming dwarfs.  If confirmed by a more detailed
ionization analysis, the optically-thick gas most likely originates in
expelled ISM gas from the nearby dwarf galaxies, either through galaxy
interactions or starburst driven outflows.

Both the LLSs toward J0357$-$4812 and J2308$-$5258 are found in a
relatively isolated environment with only one luminous, star-forming
galaxy found at $d<300$ pkpc and $|d\,v_{\rm gal}|<300$ \kms.
Incorporating results from ongoing deep- and wide-field galaxy survey
will provide a more complete view of the larger-scale galaxy
environment.  In the immediate vicinity of the LLS, however, the LLS
at $z_{\rm abs}=0.4353$ toward J0357$-$4812 is found to be associated
with a luminous galaxy of $\approx 1.4\,L_*$ at $d=67$ pkpc with the
QSO sightline intercepting the diffuse halo at $\approx 27^\circ$ from
the major axis (see \S\ 4.5).  The relative velocity offsets of the
absorbing components are consistent with the gas co-rotating with the
disk, with the high-metallicity component ($c3$) showing the largest
``lag'' in velocity.  If the low-metallicity components ($c1$ and
$c2$) originate in accreted gaseous streams that co-rotate, possible
origins for the high-metallicity component ($c3$) include fountain
flows, extraplanar gas, or disrupted satellites (see Heald \etal\ 2011
and Putman \etal\ 2012 for a review).

In contrast, the LLS at $z_{\rm abs}=0.5427$ toward J2308$-$5258 is
found to be associated with a luminous, star-forming galaxy of
$\approx 2\,L_*$ at $d=32$ pkpc with the QSO sightline intercepting
the diffuse halo at $\approx 28^\circ$ from the minor axis (see
\S\ 4.3).  The \MgII\ absorption profile is resolved into five
separate components with four satellite components at $|d\,v_c|>0$
($c1$, $c3$, $c5$, and $c6$) sharing comparable absorption strength,
while the corresponding $N(\HI)$ differ by a factor of $\approx\,6$.
Following the same argument described above, the different
$N_c(\MgII)/N_c(\HI)$ ratios suggest that the two outlying components
at $d\,v_c=-144$ and $+42$ \kms, $c1$ and $c6$, respectively, contain
the highest-metallicity gas in this LLS (panel {\it c} in Figure 13).
The combination of a close proximity to a star-forming
galaxy and a geometric alignment of the QSO sightline near the minor
axis of the galaxy makes starburst outflows an attractive scenario for
the LLS (e.g., Heckman \etal\ 1990; Shen \etal\ 2013).  However, the
large scatter in gas metallicity inferred from $N_c(\MgII)/N_c(\HI)$
between different components also suggests that additional sources are
needed to fully explain all the components seen in the LLS,
particularly the central component ($c4$) that dominates the total
$N(\HI)$ (e.g., Hafen \etal\ 2019).

In summary, our galaxy survey is sensitive for detecting galaxies
fainter than $0.1\,L_*$ at $d\apl 300$ kpc from the LLSs.  A diverse
range of galaxy properties is seen around these optically-thick
absorbers, from a low-mass dwarf galaxy pair, a co-rotating gaseous
halo/disk, a star-forming galaxy, a massive quiescent halo, to a
galaxy group.  The closest galaxies found for the LLSs have projected
distance ranging from $d=16$ pkpc to $d=72$ pkpc and intrinsic
luminosity from $\approx 0.01\,L_*$ (or $\log\,\mstar/\msun=7.8$) to
$\approx 3\,L_*$ (or $\log\,\mstar/\msun=10.9$).  This wide range of
galaxy properties uncovered for the sample of five LLSs encompasses
what is seen previously for individual systems at $z_{\rm abs}<1$
(e.g., Kacprzak \etal\ 2010; Neeleman \etal\ 2016; P\'eroux
\etal\ 2017; Chen \etal\ 2019a), highlighting the complex origins of
these absorbers.

While no clear correlation is seen between the observed overdensity of
galaxies and the kinematic spread of low-ionization species, the
observed absorption strengths of low-ionization transitions appear to
correlate with the projected distance of the closest galaxy found.
Specifically, both LLSs with weak \MgII\ features of
$\log\,N(\MgII)/\cmjj<13$ (along the sightlines toward J2135$-$5316
and J0357$-$4812) have the closest galaxy found at $d>60$ pkpc,
whereas the remaining three have $\log\,N(\MgII)/\cmjj>13.4$ and a
galaxy found at $d<50$ pkpc.  The observed trend is consistent with
previous findings that the distance to a galaxy is a more dominant
factor than the mass of the galaxies in driving the observed
\MgII\ absorption strength (e.g., Chen \etal\ 2010a,b).

%
%


\section{Summary and Conclusions}

We present initial results from the CUBS program, designed to map 
diffuse baryonic structures at $z\apl 1$, using absorption-line
spectroscopy of 15 UV bright QSOs with matching deep galaxy survey
data.  CUBS QSOs are selected based on their NUV brightness to avoid
biases against LLSs at $z_{\rm abs}<1$, which are expected to
attenuate the background QSO light in the FUV spectral window.
Combining high-quality {\it HST} COS FUV absorption spectra with
optical echelle spectra of the QSOs and deep galaxy survey data of the
QSO fields obtained from the ground, we carry out a joint study of LLSs
and their associated galaxies to gain insight into the connection
between optically-thick gas and galaxies, as well as the physical
origin of metal-line absorbers.  Our findings are summarized below:

(1) The COS spectra of the 15 NUV bright QSOs in the CUBS sample
provide a total redshift survey pathlength of $\Delta\,z_{LL}=9.3$ for
searching for LLSs.  We report five new LLSs of
$\log\,N(\HI)/\cmjj\apg 17.2$ and five new pLLS of
$\log\,N(\HI)/\cmjj=16.5-17.2$, leading to a number density of
$n(z)=0.43_{-0.18}^{+0.26}$ for LLS and $n(z)=1.08_{-0.25}^{+0.31}$
for pLLS and LLS combined at $z_{\rm abs} < 1$.  While the
uncertainties are large due to the small QSO sample, there is a
tantalizing indication that previously reported $n(z)$ of these strong
\HI\ absorbers may have been underestimated, possibly due to FUV
selection bias.

(2) All newly uncovered LLSs exhibit a multi-component structure and
associated heavy ions from low-, intermediate-, to high-ionization
states.  There is broad agreement in the overall line profiles of
both low- and intermediate-ionization species, indicating that
different ionic species are linked kinematically and share a common
velocity field.  At the same time, \OVI\ is detected in all but the
DLA at $z_{\rm abs}=0.5762$ toward J0111$-$0316.  Three of the four
\OVI-bearing optically-thick absorbers display a broad and relatively
shallow absorption profile that traces the broad kinematic
properties of associated low- and intermediate-ionization species,
while the 4th LLS displays only a single strong \OVI\ in one 
\HI\ component that contains $\approx 0.2$\% of the total $N(\HI)$.

(3) Large differential $N(\MgII)/N(\HI)$ ratios, by more than a factor
of 10, are directly seen across different components of individual
LLSs, suggesting inhomogeneous chemical enrichment as well as
ionization state in individual galaxy halos.  A detailed ionization
analysis presented in Zahedy \etal\ (2020) will quantify the extent of
density and metallicity fluctuations in individual systems.  Here we
show that high $N(\MgII)/N(\HI)$ components would be missed in integrated
column density calculations, which underscores the scientific value of
high-resolution absorption spectra in resolving the clumpy, multiphase
gas.

(4) MUSE integral field spectroscopic data, covering a $1'\times 1'$
field of view, has provided an ultra-deep view of the environment
around three of the five new LLSs.  The galaxy survey in each field is
100\% complete to $AB(r)=23$ mag (corresponding to $0.3\,L_*$ or
fainter at the redshifts of the LLSs), and between 75\% and 90\%
complete to $AB(r)=25$ mag.  A diverse range of galaxy environments is
seen around these LLSs, from a massive quiescent halo, a low-mass
dwarf galaxy pair, a co-rotating gaseous halo/disk, to a massive
galaxy group.  The closest galaxies found for the LLSs have projected
distance ranging from $d=16$ to $d=72$ pkpc and intrinsic luminosity
from $\approx 0.01\,L_*$ to $\approx 3\,L_*$.  The wide range of
galaxy environments further underscores the complexity in connecting
gas to galaxies and the need for deep galaxy survey data to fully
understand the origin of absorption-line systems.

In conclusion, we are constructing a legacy data set through the CUBS
program to enable comprehensive studies of the co-evolution of
galaxies and their surrounding gas at $z\apl 1$.  While the combined
space-based UV and ground-based optical echelle absorption
spectroscopy of distant QSOs provides a powerful tool for resolving
the ionization states and chemical enrichment histories of the diffuse
CGM/IGM over a broad range in gas density along 15 independent
sightlines, the accompanying deep and wide galaxy survey data provide
a sensitive record of the galaxy environment on both large and small
scales.  In addition to detailed ionization analyses of individual
high-$N(\HI)$ absorbers presented in Boettcher \etal\ (2020), Cooper
\etal\ (2020), Johnson \etal\ (2020), and Zahedy \etal\ (2020), in
future papers we will probe the physical conditions, chemical
composition, and kinematics of halo gas (at $d \apl 300$ pkpc) around
galaxies with diverse star formation histories and examine the
physical origin of different absorption-line systems based on their
large-scale ($\approx 1$--10 pMpc) clustering amplitudes with known
galaxies.  Together with published samples at low and high redshift,
we will also quantify the evolution of the CGM across 10 Gyrs of
cosmic evolution, with the CUBS data charting the critical period when
the star formation history of the universe underwent its precipitous
decline.

\section*{Acknowledgments}

We thank an anonymous referee for timely and constructive comments
that helped improve the presentation of the paper.  We thank Ruari
Mackenzie for prompt and helpful advice and guidance on MUSE data
reduction, and Matteo Fossati for sharing the KUBEVIZ code and
assistance on analyzing the MUSE data cubes.  We thank Dan Kelson for
his expert assistance on reducing the galaxy survey data from the
Magellan Telescopes.  HWC, EB, and MCC acknowledge partial support
from HST-GO-15163.001A and NSF AST-1715692 grants.  TC and GCR
acknowledge support from HST-GO-15163.015A.  SC gratefully
acknowledges support from Swiss National Science Foundation grant
PP00P2\_163824.  SDJ acknowledges support from a NASA Hubble
Fellowship (HST-HF2-51375.001-A).  KLC acknowledges partial support
from NSF AST-1615296.  CAFG was supported by NSF through grants
AST-1517491, AST-1715216, and CAREER award AST-1652522, by NASA
through grant 17-ATP17-0067, by STScI through grants HST-GO-14681.011,
HST-GO-14268.022-A, and HST-AR-14293.001-A, and by a Cottrell Scholar
Award from the Research Corporation for Science Advancement.  SL was
funded by project FONDECYT 1191232. This work is based on observations
made with ESO Telescopes at the Paranal Observatory under programme ID
0104.A-0147(A), observations made with the 6.5m Magellan Telescopes
located at Las Campanas Observatory, and spectroscopic data gathered
under the HST-GO-15163.01A program using the NASA/ESA Hubble Space
Telescope operated by the Space Telescope Science Institute and the
Association of Universities for Research in Astronomy, Inc., under
NASA contract NAS 5-26555.  This research has made use of NASA's
Astrophysics Data System and the NASA/IPAC Extragalactic Database
(NED) which is operated by the Jet Propulsion Laboratory, California
Institute of Technology, under contract with the National Aeronautics
and Space Administration.

\section*{Data Availability}

The data underlying this article will be shared on reasonable request to the corresponding author.

\appendix

\section{Galaxy properties in the massive group at $\bmath{d_{\rm group}=177}$ pkpc from the LLS at $\bmath{z_{\rm abs}=0.6226}$ toward J2135$-$5316}

Optical spectra of the members of the LLS-associated galaxy group
reported in \S\ 4.2.2 clearly show a wide range in the star formation
histories among the members of the galaxy group, from young
star-forming to old and evolved (Figure 6).  At $z=0.6$, MUSE does not
provide the spectral coverage necessary for observing H$\alpha$ or
[\NII].  While the spectra cover higher-order Balmer transitions and
other nebular lines, only [\OII] emission is consistently seen among
all 10 group galaxies with the \CaII\ H\&K absorption doublet and
G-band absorption being the predominant features in five group
members.  In particular, the most luminous member of the group at
$d=244$ pkpc and $d\,v_{\rm gal}=-499$ \kms\ also displays a strong
Balmer absorption series in addition to \CaII\ absorption (panel {\it
  i} in Figure 6), characteristic of a post-starburst phase (see
French \etal\ 2015 and Rowlands \etal\ 2015 for recent references).
It also shows [\NeIII] and an [\OIII]/H$\beta$ line ratio that
suggests the presence of an active galactic nucleus (AGN).

Because of a lack of robust constraints for the dust content, we
estimate an unobscured SFR based on the integrated [\OII] line flux
for each galaxy under the assumption that the presence of the [\OII]
lines is driven by the radiation field from young stars.  To determine
an integrated [\OII] line flux and the velocity offset of each pixel,
we employ a custom IDL code KUBEVIZ, kindly shared with us by
M.\ Fossati, to fit the emission doublet (see Fossati \etal\ 2016 for
a detailed description of the code).  We first smooth the data cube
using a $3\times 3$ box in the image plane (corresponding to the size
of the PSF) to improve the signal-to-noise ($S/N$) per pixel without
degrading the spatial resolution of the data.  Then for the spectrum
from each pixel, we fit the [\OII] doublet using a double Gaussian
function.  KUBEVIZ takes into account the associated error for each
spaxel in the fitting routine in order to suppress contributions from
features due to sky subtraction residuals, and delivers the best-fit
integrated line flux and associated error of the doublet, along with
the best-fit velocity and velocity dispersion maps.  We visually
inspect the fitting results across the full field and repeat the
fitting procedure as needed after modifying the input parameters.  The
total integrated [\OII] line fluxes of the group galaxies range from
$f_{\rm [O{\scriptsize II}]}=(5\pm 1)\times
10^{-19}$\,erg\,s$^{-1}$\,cm$^{-2}$ to $f_{\rm [O{\scriptsize
      II}]}=(1.52\pm 0.07)\times 10^{-16}$\,erg\,s$^{-1}$\,cm$^{-2}$,
leading to an unobscured SFR of $\approx 0.01-1.6\,\msun\,{\rm
  yr}^{-1}$ based on the star formation calibrator of Kewley
\etal\ (2004).  The line-of-sight velocity map of the galaxy group is
presented in the right panel of Figure 5.

An interesting feature of the spectra displayed in Figure 5 is the
contrast between ordinary continuum morphologies of the group galaxies
in the pseudo $r$-band image and the irregular morphology of [\OII]
line emission around two massive group members at $d\,v_{\rm
  gal}=-129$ and $-499$ \kms\ to the southwest of the LLS.  The
extended [\OII] emission morphologies of the two galaxies separated by
$\approx 20$ pkpc in projected distance and the large line-of-sight
velocity spread from $d\,v\approx -500$ \kms\ to $d\,v\approx +100$
\kms\ imply that violent interactions may be taking place between the
two galaxies, which may also be responsible for triggering the recent
episode of star formation and possible AGN phase in the post starburst
galaxy at $d\,v_{\rm gal}=-499$ \kms\ and $d=244$ pkpc (see e.g.,
Johnson \etal\ 2018).

\section{Low-mass dwarfs at $d\approx 26$ pkpc from the LLS at $z_{\rm abs}=0.3640$ toward J0248$-$4048}

The three low-mass galaxies found in the vicinity of this LLS exhibit
prominent nebular emission lines that are typical of star-forming
regions (see \S\ 4.4.2 and Figure 9).  To systematically search for
emission features associated with the LLS and to determine the
velocity offsets, we use the KUBEVIZ software to fit line-emitting
features at the redshift of the LLS in the full field covered by MUSE.
We first smooth the data cube using a $3\times 3$ box in the image
plane.
Then for the spectrum from each pixel, we fit
all available strong emission lines (in this case, the [\OII] doublet,
H$\beta$, [$\OIII$]$\lambda\lambda\,4960,5008$, and H$\alpha$; see
Figure 9 and next paragraph) simultaneously using a Gaussian function
per line and adopting a common systemic redshift and velocity width
across all lines considered.
KUBEVIZ outputs the best-fit integrated line flux and associated error
for each line, along with the best-fit velocity and velocity
dispersion maps.  Finally, we visually inspect the fitting results
across the full field and repeat the fitting procedure as needed after
modifying the input parameters.  The line-of-sight velocity map of the
three LLS-associated galaxies is presented in the right panel of
Figure 8.  No extended line emission is detected much beyond the
optical extent of the three galaxies.

The presence of prominent nebular lines in all three galaxies enable a
detailed analysis of the physical properties of these LLS-associated
galaxies, including the star formation rate (SFR), ionization
condition, ISM gas-phase metallicity, and dust content.  We first
estimate an unobscured SFR based on the total integrated H$\alpha$
line flux ($f_{\rm H\alpha}$) in the MUSE data using the conversion
from Kennicutt \& Evans (2012), ${\rm SFR}=5.37\times 10^{-42}\,L_{\rm
  H\alpha}\,\msun\,{\rm yr}^{-1}$, which is based on a Chabrier (2003)
stellar initial mass function.  The galaxies have $f_{\rm H\alpha}$
ranging from $(6.3\pm 0.2)\times 10^{-18}$ to $(4.41\pm 0.03)\times
10^{-17}$\,erg\,s$^{-1}$\,cm$^{-2}$, leading to an estimated SFR
ranging from to 0.01 to 0.11 $\msun\,{\rm yr}^{-1}$ (see columns 9 and
10 of Table 7).

Next, we examine the ionization condition of the gas by comparing the
strong line ratios, [\OIII]/H$\beta$ versus [\NII]/H$\alpha$.  None of
the three galaxies exhibit a significant [\NII] line, placing a
2-$\sigma$ upper limit on the $N2$ index, $N2\equiv\log\,{\rm
  [\NII]}\,\lambda\,6585/{\rm H}\alpha$, of $N2<-1$.  In addition,
the [\OIII]/H$\beta$ ratio of these galaxies ranges between
$\log\,{\rm [\OIII]/H}\beta=0.48$ and 0.52.  Together, the observed
[\OIII]/H$\beta$ and [\NII]/H$\alpha$ ratios indicate that the ISM is
ionized predominantly by young stars (e.g., Baldwin \etal\ 1981),
rather than by active galactic nuclei (AGN).

Finally, we estimate the ISM gas-phase metallicity and dust content
using common emission line calibrators.  We first infer a 2-$\sigma$
upper limit to the ISM gas-phase metallicity of $12 + \log({\rm
  O}/{\rm H}) < 8.3$ based on the absence of [\NII] and the $N2$
calibration of Marino \etal\ (2013), $12 + \log({\rm O}/{\rm H}) =
8.74 + 0.46\times N2$.  Then we compute the $R_{23}$ index, defined as
$R_{23}\equiv$\,($[\OII]\,\lambda\lambda\,3726,3729+[\OIII]\,\lambda\lambda\,4960,5008$)/H$\beta$.
For an accurate estimate of $R_{23}$, we assess the amount of dust
extinction correction using the observed H$\alpha$/H$\beta$ flux
ratio.  Following the prescription described in Calzetti
\etal\ (2000)\footnote{Adopting the extinction law for the Small
  Magellanic Cloud from Gordon \etal\ (2003) does not change the
  extinction-corrected line ratios significantly.}, we estimate the
color excess $E(B-V)$ according to $E(B-V)=1.97\,\log\,([f_{\rm
    H\alpha}/f_{\rm H\beta}]/2.86)$, and find $E(B-V)=0.15\pm 0.01$,
$0.23\pm 0.04$, and $0.07\pm 0.02$ for the galaxies at $d=15$, 37, and
150 pkpc, respectively.  The wavelength-dependent extinction magnitude
$A(\lambda)$ is related to $E(B-V)$ following
$A(\lambda)=k(\lambda)E(B-V)$, where $k(\lambda)$ is the dust
extinction law.  Adopting the Calzetti (1997) extinction law, $k({\rm
  [\OII]})=5.86$, $k({\rm H\beta})=4.60$, $k({\rm [\OIII]})=4.46$, and
$k({\rm H\alpha})=3.33$.  This exercise enables us to estimate the
oxygen abundance using the $R_{23}$ index based on
extinction-corrected line ratios.  Based on the calibration of Yin
\etal\ (2007), $12 + \log({\rm O}/{\rm H}) = 6.486 + 1.401\times
\log\,R_{23}$, which is justified by the upper limit of the $N2$
index, we find $12 + \log({\rm O}/{\rm H}) = 7.7$, 7.5, and
7.6\footnote{For comparison, the Sun has $12 + \log({\rm O}/{\rm
    H})_\odot = 8.69\pm 0.05$ (Asplund \etal\ 2009).} for the galaxies
with increasing $d$ (see column 11 of Table 7).  Uncertainties in the
gas phase metallicity are driven by the systematic uncertainty of the
$R_{23}$ calibration and it is estimated to be 0.1 dex (Yin
\etal\ 2007).

While robust emission-line fluxes have been obtained for all three
dwarf galaxies, accurate broad-band photometry is only feasible for
the two galaxies at $d=37$ and 150 pkpc, not affected by the glare of
the QSO.  The observed $r$-band magnitudes of these two galaxies are
$AB(r)=25.1$ and 24.1, respectively.  We estimate an intrinsic
$r$-band absolute magnitude $M_r$ using a star-forming galaxy template
and find $M_r=-16.3$ and $-17.3$ for the two galaxies, corresponding
to $0.01$ and $0.025\,L_*$ adopting $M_{r_*}=-21.3$ for blue galaxies
from Cool \etal\ (2012).  We further estimate the underlying stellar
mass \mstar\ using the prescription for blue galaxies presented in
Johnson \etal\ (2015), and find $\log\,\mstar/\msun=7.8$ and 8.2,
respectively.  Both galaxies are found to be exceedingly faint and
low-mass.  Based on the estimated ISM gas phase metallicity of
$\approx 10$\% solar and adopting the mass-metallicity relation of
dwarf galaxies (e.g., Berg \etal\ 2012), we argue that the closest
galaxy at $d=15$ pkpc is also likely to be a low-mass dwarf of
$\log\,\mstar/\msun\approx 8$.
The observed and derived properties of
these three galaxies
are summarized in columns (2) through (11) of Table 7.

\section{Properties of a rotating disk galaxy at $d=67$ pkpc from the LLS at $z_{\rm abs}=0.4353$ toward J0357$-$4812}

The optical spectrum of the absorbing galaxy is characterized by
strong absorption features, together with weak [\OII] and H$\beta$
emission lines, indicating that this is an early-type galaxy (see
\S\ 4.5.2 and Figure 11).  At $z=0.4353$, MUSE does not provide the
spectral coverage necessary for observing H$\alpha$, limiting our
ability in obtaining robust constraints for the dust content, and
consequently for the ISM gas-phase metallicity and ionization
condition of the galaxy.  However, the presence of the [\OII] emission
doublet does enable measurements of the velocity field across the
galaxy as well as an estimate of the unobscured SFR.  To determine an
integrated [\OII] line flux and the velocity centroid at each pixel,
we fit the emission doublet using KUBEVIZ, visually inspect the
fitting results across the full field, and repeat the fitting
procedure as needed after modifying the input parameters.  We find a
total integrated [\OII] line flux over the entire galaxy of $f_{\rm
  [O{\scriptsize II}]}=(6.14\pm 0.06)\times
10^{-17}$\,erg\,s$^{-1}$\,cm$^{-2}$, leading to an unobscured SFR of
$\approx 0.28\,\msun\,{\rm yr}^{-1}$ based on the star formation
calibrator of Kewley \etal\ (2004).

The observed velocity gradient along the long axis in panel ({\it b})
of Figure 10 supports the presence of a rotating disk. To investigate
the relative motion of the LLS with respect to the rotation of the
galaxy, we first estimate the inclination and orientation of the
underlying disk using the pseudo $r$-band image.  The outline of the
galaxy appears to be largely well-defined by an ellipse in the pseudo
$r$-band image with a mild irregular feature toward the northwest
corner, as suggested also by the [\OII] emitting morphology.  We find
that the galaxy can be characterized by an inclination angle of
$i\approx 44^\circ$ and a position angle of the major axis of
$\alpha\approx 120^\circ$, north through east.  If there exists an
extended gaseous disk, then the QSO sightline probes the gas at
$\approx 27^\circ$ from the major axis.  Next, we de-project both the
projected distance $d$ and $d\,v_{\rm gal}$ observed along the major
axis onto the disk plane.  Following the prescription described in
Chen \etal\ (2005), the galactocentric radius $R$ is related to $d$
according to $R=d\,\sqrt{1+\sin^2(\phi-\alpha)\tan^2(i)}$, while the
de-projected velocity is related to the observed line-of-sight
velocity according to $\Delta\,v_{\rm
  deproj}=\Delta\,v\,\sqrt{1+\sin^2(\phi-\alpha)\tan^2(i)}/\cos(\phi-\alpha)/\sin(i)$,
where $\phi$ is the position angle of the slit.  For the galaxy disk,
the MUSE data cube enables us to place a pseudo slit along the major
axis, in which case $\phi_{\rm gal}=\alpha$, whereas for the LLS
$\phi$ is dictated by the location of the QSO probe relative to the
galaxy and we find $\phi_{\rm QSO}\approx 93^\circ$.
  
The de-projected rotation velocity along the disk plane as a function
of galactocentric radius in panel ({\it c}) of Figure 10 also includes
the three strongest \HI\ absorbing components (c1, c2, and c3 from
Table 6) of the LLS along the QSO sightline for comparison (open
circles in panel {\it c} of Figure 10).  Under the assumption that the
gas is moving along the plane extended from the inclined optical disk,
the zero velocity in panel ({\it c}) of Figure 10 corresponds to the
systemic redshift of the galaxy.  While the rotation curve appears to
flatten at $d\,v_{\rm deproj}\approx +225$ \kms\ on the east side of
the galaxy, it continues to extend beyond $|d\,v_{\rm deproj}|\approx
300$ \kms\ on the west side which is likely related to the extended
structure revealed in [\OII] emission.

\label{lastpage}

\end{document}